\documentclass[a4paper,american,cleveref, autoref, thm-restate]{lipics-v2021}

\usepackage{graphicx}
\usepackage{color,soul}

\usepackage{amssymb,amsmath,amsthm,ragged2e}

\usepackage{caption,subcaption}

\usepackage{algorithm}
\usepackage[]{algpseudocode}
\usepackage{tikz}
\usepackage{circuitikz}
\usepackage{pgfplots}
\usepackage{verbatim}
\usepackage{mathrsfs}
\usepackage{lineno}
\usetikzlibrary{arrows}

\usepackage{enumerate}

\usepackage{cleveref}
\usepackage[]{cite}
\usetikzlibrary{positioning,decorations.pathmorphing,calc}

\graphicspath{{figs/}}

\pgfplotsset{compat=1.15}

\usepackage{xcolor}
\usepackage{soul}

\providecommand{\wbdpercent}{10\%}

\Crefname{table}{Table}{Tables}
\Crefname{obs}{Observation}{Observations}
\Crefname{cor}{Corollary}{Corollaries}
\Crefname{algorithm}{Algorithm}{Algorithms}
\Crefname{lemma}{Lemma}{Lemmas}

\algnewcommand\algorithmicforeach{\textbf{for each}}
\algdef{S}[FOR]{ForEach}[1]{\algorithmicforeach\ #1\ \algorithmicdo}

\newcommand{\REM}[1]{}

\providecommand{\carin}{car}
\providecommand{\carins}{cars}
\providecommand{\vin}{order}
\providecommand{\vins}{orders}
\providecommand{\blendnumber}{sequence number}

\providecommand{\blendnumbers}{sequence numbers}

\providecommand{\bodybuffer}{Body Buffer}

\author{Andreas Karrenbauer}{Max Planck Institute for Informatics, Saarland Informatics Campus}{}{}{}
\author{Bernd Kuhn}{Ford-Werke GmbH, Saarlouis, Germany}{}{}{}
\author{Kurt Mehlhorn}{Max Planck Institute for Informatics, Saarland Informatics Campus}{}{}{}
\author{Paolo Luigi Rinaldi}{Max Planck Institute for Informatics, Saarland Informatics Campus}{}{}{}

\authorrunning{$ $}

\Copyright{}
\hideLIPIcs
\nolinenumbers

\begin{document}

\date{}
\title{Optimizing Car Resequencing on Mixed-Model Assembly Lines: Algorithm Development and Deployment}
\titlerunning{Optimizing Car Resequencing on Mixed-Model Assembly Lines}

\maketitle


\begin{abstract}

The mixed-model assembly line (MMAL) is a production system used in the automobile industry to manufacture different car models on the same conveyor, offering a high degree of product customization and flexibility. However, the MMAL also poses challenges, such as finding optimal sequences of models satisfying multiple constraints and objectives related to production performance, quality, and delivery -- including minimizing the number of color changeovers in the Paint Shop, balancing the workload and setup times on the assembly line, and meeting  customer demand and delivery deadlines. We propose a multi-objective algorithm to solve the MMAL resequencing problem under consideration of all these aspects simultaneously. We also present empirical results obtained from recorded event data of the production process over $4$ weeks following the deployment of our algorithm in the Saarlouis plant of Ford-Werke GmbH. We achieved an improvement of the average batch size of about $30\%$ over the old control software translating to a $23\%$ reduction of color changeovers. Moreover, we reduced the spread of cars planned for a specific date by $10\%$, reducing the risk of delays in delivery. We discuss effectiveness and robustness of our algorithm in improving production performance and quality as well as trade-offs and limitations.
\end{abstract}

\newpage
\section{Introduction}

In general, car factories are comprised of three sections: the Body Shop, the Paint Shop, and the Final Assembly. In the Body Shop the chassis are built, in the Paint Shop the chassis are painted; and in the Final Assembly all the components (engine, breaks, interior, \ldots) are assembled to create a complete car, see Figure~\ref{fig:wherearewe}.

\newcommand{\mybuffer}[5]{%
    \pgfmathsetmacro{\numlines}{#1}
    \pgfmathsetmacro{\maxheight}{#2}
    \pgfmathsetmacro{\xleft}{#3} %
    \pgfmathsetmacro{\xright}{#4} %
    \pgfmathsetmacro{\yverticalsegments}{#5}
    \pgfmathsetmacro{\totlen}{\xright - \xleft} %

    \def\ratioSegment{1}
    \def\ratioCurve{2}
    \def\ratioLines{4}
    \def\totalRatio{10}

    \pgfmathsetmacro{\xLeftSegmentEnd}{\xleft + \ratioSegment * \totlen / \totalRatio}
    \pgfmathsetmacro{\xLeftCurveStart}{\xLeftSegmentEnd}
    \pgfmathsetmacro{\xParallelLinesStart}{\xleft + (\ratioSegment + \ratioCurve) * \totlen / \totalRatio}
    \pgfmathsetmacro{\xParallelLinesEnd}{\xleft + (\ratioSegment + \ratioCurve + \ratioLines) * \totlen / \totalRatio}
    \pgfmathsetmacro{\xRightCurveStart}{\xParallelLinesEnd}
    \pgfmathsetmacro{\xRightSegmentStart}{\xleft + (\ratioSegment + \ratioCurve + \ratioLines + \ratioCurve) * \totlen / \totalRatio}
    \pgfmathsetmacro{\xRightSegmentEnd}{\xright}

    \ifnum\numlines>1
        \pgfmathsetmacro{\linesep}{\maxheight/(\numlines-1)}
    \else
        \def\linesep{0}
    \fi

    \draw (\xleft,\yverticalsegments) -- (\xLeftSegmentEnd,\yverticalsegments) coordinate (left_vert);
    \draw (\xRightSegmentStart,\yverticalsegments) -- (\xRightSegmentEnd,\yverticalsegments) coordinate (right_vert);

    \foreach \i in {1,...,\numlines}
    {
        \pgfmathsetmacro{\currentshift}{\yverticalsegments - \maxheight/2 + (\i-1) * \linesep}
        \draw (\xParallelLinesStart,{\currentshift}) -- (\xParallelLinesEnd,{\currentshift}) coordinate (hline-\i-right);
        \draw (\xLeftSegmentEnd,\yverticalsegments) .. controls (\xLeftCurveStart + 0.5*\ratioCurve * \totlen / \totalRatio,\yverticalsegments) and (\xParallelLinesStart - 0.5*\ratioCurve * \totlen / \totalRatio,{\currentshift}) .. (\xParallelLinesStart,{\currentshift});
        \draw (\xParallelLinesEnd,{\currentshift}) .. controls (\xParallelLinesEnd + 0.5*\ratioCurve * \totlen / \totalRatio,\currentshift) and (\xRightSegmentStart - 0.5*\ratioCurve * \totlen / \totalRatio,\yverticalsegments) .. (\xRightSegmentStart,\yverticalsegments);
    }
}

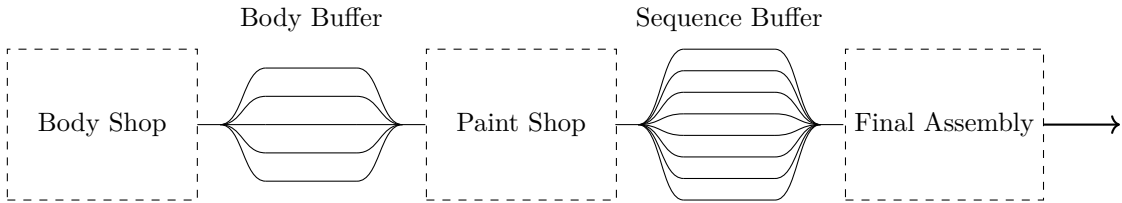
\begin{figure}[htb!]
    \centering
\begin{tikzpicture}
    \node[draw, dashed, rectangle, minimum width=2.5cm, minimum height=2cm] (body_buffer) {Body Shop};
    
    \mybuffer{5}{1.5}{1.25}{4.25}{0}  %
    \node[align=center] at (2.75, 1.4) {\bodybuffer{}};
    
    \node[draw, dashed, rectangle, minimum width=2.5cm, minimum height=2cm, right=3cm of body_buffer] (paint_shop) {Paint Shop};
    
    \mybuffer{8}{2}{6.75}{9.75}{0}   %
    \node[align=center] at (8.25, 1.4) {Sequence Buffer};
    
    \node[draw, dashed, rectangle, minimum width=2cm, minimum height=2cm, right=3cm of paint_shop] (final_assembly) {Final Assembly};

    \draw[thick,->] (final_assembly.east) -- ++(1cm,0);

\end{tikzpicture}
    \caption{A schematic representation of a car factory. In the Saarlouis plant, the \bodybuffer{} (left) has $13$ lanes and a total capacity of $148$ cars, while the Sequence Buffer (right) has $21$ lanes and a total capacity of $250$ cars.}
    \label{fig:wherearewe}
\end{figure}

The sections are separated by buffers. The buffers dampen the propagation of disruptions and allow the reordering of cars in order to optimize production. The buffers typically consist of several parallel first-in-first-out lanes, see Figure~\ref{fig:wherearewe}. Sometimes, random access stores are used. For this paper, buffers are assumed to consist of parallel FIFO lanes. In the literature, they are also called mix banks, or selectivity banks~\cite{decomposition_boysen_2013}.

Cars come in different models, even cars produced in the same factory come in many different models: limousine or station wagon, four doors or five doors, sunroof, engine type, color, \ldots. The number of different models goes into the millions~\cite{meyr2009supply}. The fabrication process in a car factory is therefore organized as a so-called mixed-model assembly line (MMAL). It allows the production of different car models on a single production line. However, it also poses a major challenge. Suppose the set of cars to be produced in a certain period has been fixed; in what sequence
should the cars be built? The sequence has a considerable impact on costs, production flow, the environment, and other factors. In the Paint Shop, color changes are to be minimized, as each color change requires cleaning the painting equipment and negatively affects the environment, production efficiency, and quality. In the Final Assembly, different features may require different installation times, e.g., a more advanced sound system may require more time than a simple sound system. Time consuming features must be spread out over the sequence so as to guarantee smooth operation in the Final Assembly. The production sequence should therefore satisfy a number of constraints (more on constraints in Section~\ref{constraints}). Constraint violations may delay production or may even result in a stoppage of the assembly line. On top of the constraints, during assembly, the supplies for the assembly process are often provided just-in-time. The supplies are guaranteed to be available for the cars, assuming the original sequence. The position of the cars in the sequence is encoded in a number that we call \blendnumber{}. It is important to generate a sequence as close as possible to the original distribution in order to maximize the chances of having the supplies ready for every car during assembly. The sequencing must not only take into account the objectives of the plant, but also the expectations of the customers and the market. For example, an excessive delay in the delivery of a car will negatively affect customer satisfaction and may reduce future demand. It should be noted that the different objectives of what constitutes a good sequence are conflicting and that the optimality criteria are different for the different stages of the plant. The buffers between the stages can therefore be used for \emph{resequencing}.

We propose an algorithm that addresses the resequencing problem for the \bodybuffer{} between Body Shop and Paint Shop by considering the constraints of the plant, the requirements of the Paint Shop, and the delivery deadlines for the cars. We also present the results of applying our algorithm in the Saarlouis plant of Ford-Werke GmbH. 
We show that, with our algorithm, the Average Batch Size (ABS) of the colors (see Section~\ref{subsec:abs} for details) increases by about $30\%$, which translates to a reduction of the color changeovers by about $23\%$.
This result has relevant and direct impact on the production costs, workload, and environment. We also managed to reduce the spread factor of the cars planned for a specific date by $10\%$. With cars spread over a more narrow section of the sequence, it is more likely that they will be delivered on the same day. 
On top of improving these Key Performance Indicators (KPIs), our sequences satisfy crucial constraints set by the plant operators, and avoid slowdowns or stoppages of the assembly line.

Section~\ref{related work} discusses related work. In Section~\ref{setting}, we explain the setting where we operate, which can be adapted to different settings in different plants as well. Section~\ref{algorithm} describes the algorithm adopted, mentioning other versions of the algorithm that we also considered. The results are presented in Section~\ref{results}, while in Section~\ref{conclusion} we conclude the paper with a general discussion. The reader may proceed directly to the results section.

\section{Related Work}\label{related work}

There is a significant amount of literature on the MMAL sequencing and resequencing problem
\cite{colourbatching_sun_2015,decomposition_boysen_2013,resequencing_boysen_2012,environmentaware_zhang_2017,multiobjective_leng_2022,deep_khadivi_2023,hybrid_leng_2019,paint_ko_2016,virtual_xu_2016,Huang24092024,SUN2024109990}. The papers discuss different aspects of the problem, use different algorithmic techniques, and aim to produce either optimal results or approximations. 

Earlier studies \cite{colourbatching_sun_2015, decomposition_boysen_2013, virtual_xu_2016, PONTES2024} tend to focus on a single objective, such as minimizing the number of color changes, or the constraint (or rule) violations.
However, these studies are often inadequate for practical applications, as they ignore many other important aspects that affect production performance and quality. Therefore, more recent research has shifted to addressing the resequencing problem with multiple objectives, such as balancing the workload, reducing the environmental impact, satisfying the customer demand, and so on \cite{bilevel_chen_2021,environmentaware_zhang_2017,multiobjective_leng_2022}. However, there is still a lack of studies that demonstrate the practical implementation and effectiveness of these multi-objective methods, as most of them rely on simulated data and experiments.

\section{Setting and Algorithm}\label{SnA}
In the following sections, we introduce the Key Performance Indicators (KPI) and the Algorithm. With the former, we define our goals, and with the latter, we describe the settings in detail and show how we address these goals algorithmically.

Key Performance Indicators are discussed in Section~\ref{subsec:kpi}, and the Algorithm is discussed in Section~\ref{algorithm}.

\subsection{Key Performance Indicators} \label{subsec:kpi}
We now describe the KPIs for our work, namely the Average Batch Size, the Constraints, and the Built-to-Date.

\subsubsection{Average Batch Size} \label{subsec:abs}
A good sequence for the Paint Shop is one that results in the least possible number of color changeovers. By producing sequences that minimize the number of color changeovers during the painting process, we can reduce costs for rinsing the equipment by minimizing the number of times that two consecutive cars have different colors. Equivalently, we aim to maximize the \emph{Average Batch Size} (ABS), which is the average number of consecutive cars with the same color during the actual painting.

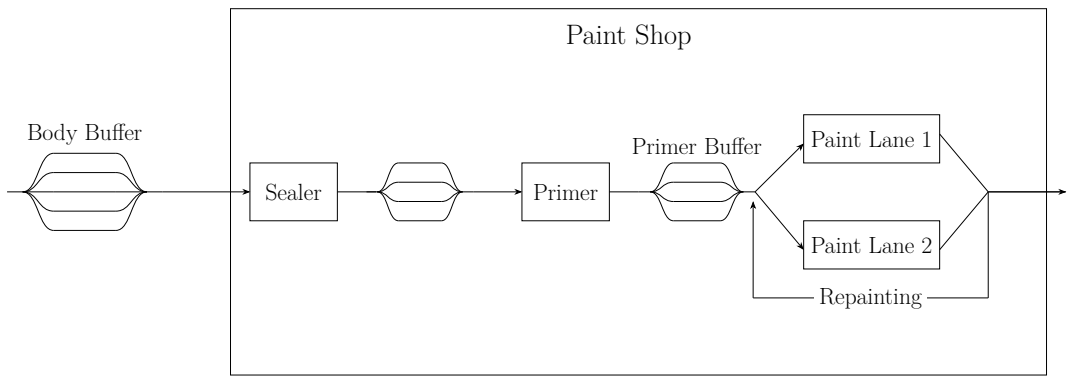
\begin{figure}[htb!]
    \centering
\resizebox{1\textwidth}{!}{%
\begin{circuitikz}
\tikzstyle{every node}=[font=\LARGE]
\mybuffer{5}{2}{1.25}{5.25}{12}  %
\node[align=center] at (3.25, 13.5) {\bodybuffer{}};
\draw [ line width=0.5pt ] (7.5,12.75) rectangle  node {\LARGE Sealer} (9.75,11.25);
\draw [line width=0.5pt, -, >=Stealth] (9.75,12) -- (10.75,12);
\mybuffer{4}{1.5}{10.5}{13.25}{12}   %
\draw [line width=0.5pt, ->, >=Stealth] (13.25,12) -- (14.5,12);
\draw [ line width=0.5pt ] (14.5,12.75) rectangle  node {\LARGE Primer} (16.75,11.25);
\draw [ line width=0.5pt ] (7,16.75) rectangle (28,7.25);
\draw [line width=0.5pt, -, >=Stealth] (16.75,12) -- (17.5,12);
\node[align=center] at (19, 13.2) {Primer Buffer};
\mybuffer{4}{1.5}{17.5}{20.5}{12}   %
\draw [ line width=0.5pt ] (21.75,14) rectangle  node {\LARGE Paint Lane 1} (25.25,12.75);
\draw [ line width=0.5pt ] (21.75,11.25) rectangle  node {\LARGE Paint Lane 2} (25.25,10);
\draw [line width=0.5pt, ->, >=Stealth] (20.5,12) -- (21.75,13.25);
\draw [line width=0.5pt, ->, >=Stealth] (20.5,12) -- (21.75,10.5);
\draw [line width=0.5pt, short] (25.25,13.5) -- (26.5,12);
\draw [line width=0.5pt, short] (25.25,10.5) -- (26.5,12);
\draw [line width=0.5pt, ->, >=Stealth] (26.5,12) -- (28.5,12);
\draw [line width=0.5pt, ->, >=Stealth] (5.25,12) -- (7.5,12);
\node [font=\huge] at (17.25,16) {Paint Shop};
\draw [line width=0.5pt, short] (26.5,12) -- (28.5,12);
\draw [line width=0.5pt, short] (26.5,12) -- (26.5,9.25);
\draw [line width=0.5pt, short] (26.5,9.25) -- (20.45,9.25)node[pos=0.5, fill=white]{Repainting};
\draw [line width=0.5pt, ->, >=Stealth] (20.45,9.25) -- (20.45,11.75);
\end{circuitikz}
}%
    \caption{Abstract Paint Shop scheme. The Primer buffer has six lanes.}
    \label{fig:paintshop_scheme}
\end{figure}

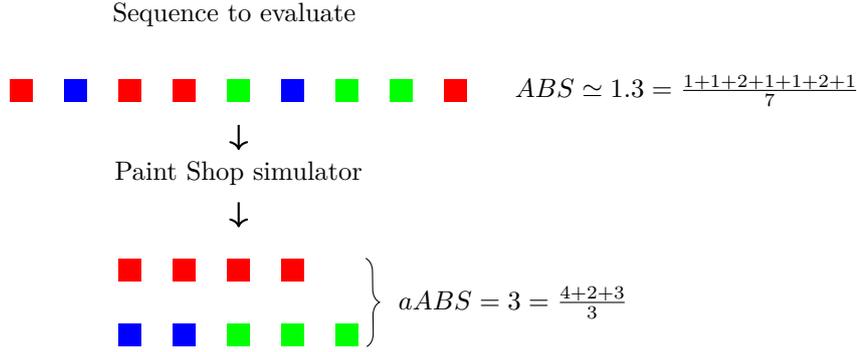
\begin{figure}[htb!]
    \centering
    \begin{tikzpicture}
    \node[above, anchor=center] (initial_text) at (2, 0) {Sequence to evaluate}; %

    \def\squareSize{0.3cm}
    \def\xShift{0.4cm}
    \def\yShift{1.0cm} %
    \def\arrowVerticalShift{0.3cm} %
    \def\horizontalListShift{0.8cm} %

    \node[fill=red, minimum size=\squareSize, inner sep=0pt] (sq1_1) at (-4*\xShift+\horizontalListShift, -1cm) {};
    \node[fill=blue, minimum size=\squareSize, inner sep=0pt, right=\xShift of sq1_1] (sq1_2) {};
    \node[fill=red, minimum size=\squareSize, inner sep=0pt, right=\xShift of sq1_2] (sq1_3) {};
    \node[fill=red, minimum size=\squareSize, inner sep=0pt, right=\xShift of sq1_3] (sq1_4) {};
    \node[fill=green, minimum size=\squareSize, inner sep=0pt, right=\xShift of sq1_4] (sq1_5) {};
    \node[fill=blue, minimum size=\squareSize, inner sep=0pt, right=\xShift of sq1_5] (sq1_6) {};
    \node[fill=green, minimum size=\squareSize, inner sep=0pt, right=\xShift of sq1_6] (sq1_7) {};
    \node[fill=green, minimum size=\squareSize, inner sep=0pt, right=\xShift of sq1_7] (sq1_8) {};
    \node[fill=red, minimum size=\squareSize, inner sep=0pt, right=\xShift of sq1_8] (sq1_9) {}; %

    \node[right=0.5cm] at (sq1_9.east) (abs_1_3) {$ABS \simeq 1.3 = \frac{1 + 1 + 2 + 1 + 1 + 2 + 1}{7}$};

    \draw[->, thick] ([yshift=-\arrowVerticalShift]sq1_5.south) -- ++(0, -\yShift/3); %

    \node[below=\yShift*2/3 of sq1_5] (paint_shop) {Paint Shop simulator};

    \draw[->, thick] ([yshift=-\arrowVerticalShift/3]paint_shop.south) -- ++(0, -\yShift/3); %

    \node[fill=red, minimum size=\squareSize, inner sep=0pt] (sq2_1) at (sq1_3 |- paint_shop.south -| sq1_3)  [yshift=-\yShift] {};
    \node[fill=red, minimum size=\squareSize, inner sep=0pt, right=\xShift of sq2_1] (sq2_2) {};
    \node[fill=red, minimum size=\squareSize, inner sep=0pt, right=\xShift of sq2_2] (sq2_3) {};
    \node[fill=red, minimum size=\squareSize, inner sep=0pt, right=\xShift of sq2_3] (sq2_4) {};

    \node[fill=blue, minimum size=\squareSize, inner sep=0pt] (sq2_5) at (sq2_1 |- sq2_1.south)  [yshift=-\squareSize-\xShift] {}; %
    \node[fill=blue, minimum size=\squareSize, inner sep=0pt, right=\xShift of sq2_5] (sq2_6) {};
    \node[fill=green, minimum size=\squareSize, inner sep=0pt, right=\xShift of sq2_6] (sq2_7) {};
    \node[fill=green, minimum size=\squareSize, inner sep=0pt, right=\xShift of sq2_7] (sq2_8) {};
    \node[fill=green, minimum size=\squareSize, inner sep=0pt, right=\xShift of sq2_8] (sq2_9) {}; %

    \draw [decorate,decoration={brace,amplitude=5pt,mirror}]
      ([xshift=0.1cm]sq2_9.south east) -- ([xshift=0.8cm]sq2_4.north east) node[midway, right=0.3cm] (aabs_3) {$aABS = 3 = \frac{4 + 2 + 3}{3}$};

\end{tikzpicture}
    \caption{The sequence that comes out of the \bodybuffer{} is not the sequence in which the cars will be painted because they go through smaller buffers and are then distributed over two paint lanes. However, we can simulate the behavior of that system and thereby evaluate the sequence leaving the \bodybuffer{} by looking at how it will most likely be reshuffled.}
    \label{fig:abs_aabs}
\end{figure}

There is a complication that we need to mention. In the Saarlouis plant, the \bodybuffer{} does not feed directly into the paint lanes. Rather, the Paint Shop has significant internal structure: there are four processing units, the sealer\footnote{The sealer puts on extra protection on critical parts in preparation for painting.}, the primer\footnote{A base paint is applied.}, and two parallel paint lanes\footnote{The actual paint is applied.}. The processing units are separated by small buffers, see Figure~\ref{fig:paintshop_scheme}. Also, paint is an error-prone process, and about 10\% of bodies need additional work. The buffers inside the Paint Shop are operated by legacy code, which is not fully documented and which we were not allowed to modify. We reverse-engineered the behavior of the Paint Shop and implemented a simulator for it. During the development of our algorithm, we used the simulator to assess the quality of the sequences leaving the \bodybuffer{}. The quality measure is the average batch size of the sequences entering the paint lanes. We refer to this measure as \emph{assessed Average Batch Size (aABS)} illustrated in Figure~\ref{fig:abs_aabs}. We discuss our algorithm for the \bodybuffer{} in Section~\ref{algorithm}. After the deployment of our algorithm, we verified our predictions with the actual aggregated numbers from the Paint Shop.

\subsubsection{Constraints}
\label{constraints}

Some of the features of the cars (e.g., two-tone color) have certain requirements, i.e., only a certain number of cars with that feature are to be contained in each window of a specified number of cars or in a specified period of time.
Such requirements can be specified by so-called \emph{constraints}. Each constraint has a logical formula of features in conjunctive normal form (CNF) to determine the affected orders. Some constraints are more important than others, so each one is associated with a weight. We minimize the total weight of the violated constraints to minimize the negative effects of the build sequence on the production process.

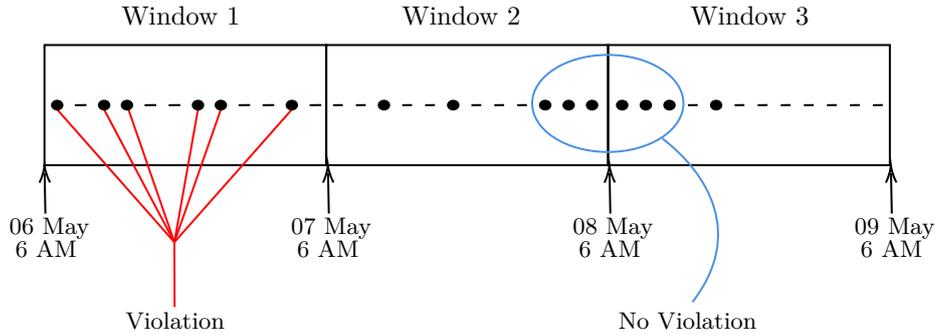
\begin{figure}[htb!]
    \centering
    
\tikzset{every picture/.style={line width=0.75pt}} %

\begin{tikzpicture}[x=1.5pt,y=1.5pt,yscale=-1,xscale=1]

\draw   (79.52,110.48) -- (149.86,110.48) -- (149.86,140.71) -- (79.52,140.71) -- cycle ;
\draw   (149.86,110.48) -- (220.19,110.48) -- (220.19,140.71) -- (149.86,140.71) -- cycle ;
\draw   (220.19,110.48) -- (290.52,110.48) -- (290.52,140.71) -- (220.19,140.71) -- cycle ;
\draw    (87.1,125.61) -- (90.01,125.58) ;
\draw    (104.52,125.61) -- (107.43,125.58) ;
\draw    (110.53,125.61) -- (113.44,125.58) ;
\draw    (127.95,125.61) -- (130.85,125.58) ;
\draw    (133.95,125.61) -- (136.86,125.58) ;
\draw    (145.69,125.61) -- (148.6,125.58) ;
\draw    (151.54,125.61) -- (154,125.58) ;
\draw    (157.25,125.61) -- (159.71,125.58) ;
\draw    (168.81,125.61) -- (171.27,125.58) ;
\draw    (174.52,125.61) -- (176.98,125.58) ;
\draw    (186.08,125.61) -- (188.54,125.58) ;
\draw    (191.79,125.61) -- (194.25,125.58) ;
\draw    (197.5,125.61) -- (199.95,125.58) ;
\draw    (240.11,125.61) -- (242.57,125.58) ;
\draw    (251.77,125.61) -- (254.23,125.58) ;
\draw    (257.53,125.61) -- (259.99,125.58) ;
\draw    (263.29,125.61) -- (265.75,125.58) ;
\draw    (269.05,125.61) -- (271.51,125.58) ;
\draw    (274.81,125.61) -- (277.26,125.58) ;
\draw    (280.56,125.61) -- (283.02,125.58) ;
\draw    (286.31,125.61) -- (288.77,125.58) ;
\draw  [fill={rgb, 255:red, 0; green, 0; blue, 0 }  ,fill opacity=1 ] (203.2,125.6) .. controls (203.2,124.99) and (203.79,124.5) .. (204.5,124.5) .. controls (205.22,124.5) and (205.81,124.99) .. (205.81,125.6) .. controls (205.81,126.2) and (205.22,126.7) .. (204.5,126.7) .. controls (203.79,126.7) and (203.2,126.2) .. (203.2,125.6) -- cycle ;
\draw  [fill={rgb, 255:red, 0; green, 0; blue, 0 }  ,fill opacity=1 ] (209.06,125.6) .. controls (209.06,124.99) and (209.64,124.5) .. (210.36,124.5) .. controls (211.08,124.5) and (211.66,124.99) .. (211.66,125.6) .. controls (211.66,126.2) and (211.08,126.7) .. (210.36,126.7) .. controls (209.64,126.7) and (209.06,126.2) .. (209.06,125.6) -- cycle ;
\draw  [fill={rgb, 255:red, 0; green, 0; blue, 0 }  ,fill opacity=1 ] (214.9,125.6) .. controls (214.9,124.99) and (215.48,124.5) .. (216.2,124.5) .. controls (216.92,124.5) and (217.5,124.99) .. (217.5,125.6) .. controls (217.5,126.2) and (216.92,126.7) .. (216.2,126.7) .. controls (215.48,126.7) and (214.9,126.2) .. (214.9,125.6) -- cycle ;
\draw  [fill={rgb, 255:red, 0; green, 0; blue, 0 }  ,fill opacity=1 ] (162.96,125.6) .. controls (162.96,124.99) and (163.54,124.5) .. (164.26,124.5) .. controls (164.98,124.5) and (165.56,124.99) .. (165.56,125.6) .. controls (165.56,126.2) and (164.98,126.7) .. (164.26,126.7) .. controls (163.54,126.7) and (162.96,126.2) .. (162.96,125.6) -- cycle ;
\draw  [fill={rgb, 255:red, 0; green, 0; blue, 0 }  ,fill opacity=1 ] (180.23,125.6) .. controls (180.23,124.99) and (180.81,124.5) .. (181.53,124.5) .. controls (182.25,124.5) and (182.83,124.99) .. (182.83,125.6) .. controls (182.83,126.2) and (182.25,126.7) .. (181.53,126.7) .. controls (180.81,126.7) and (180.23,126.2) .. (180.23,125.6) -- cycle ;
\draw  [fill={rgb, 255:red, 0; green, 0; blue, 0 }  ,fill opacity=1 ] (222.4,125.6) .. controls (222.4,124.99) and (222.99,124.5) .. (223.7,124.5) .. controls (224.42,124.5) and (225.01,124.99) .. (225.01,125.6) .. controls (225.01,126.2) and (224.42,126.7) .. (223.7,126.7) .. controls (222.99,126.7) and (222.4,126.2) .. (222.4,125.6) -- cycle ;
\draw  [fill={rgb, 255:red, 0; green, 0; blue, 0 }  ,fill opacity=1 ] (228.31,125.6) .. controls (228.31,124.99) and (228.89,124.5) .. (229.61,124.5) .. controls (230.33,124.5) and (230.91,124.99) .. (230.91,125.6) .. controls (230.91,126.2) and (230.33,126.7) .. (229.61,126.7) .. controls (228.89,126.7) and (228.31,126.2) .. (228.31,125.6) -- cycle ;
\draw  [fill={rgb, 255:red, 0; green, 0; blue, 0 }  ,fill opacity=1 ] (234.21,125.6) .. controls (234.21,124.99) and (234.79,124.5) .. (235.51,124.5) .. controls (236.23,124.5) and (236.81,124.99) .. (236.81,125.6) .. controls (236.81,126.2) and (236.23,126.7) .. (235.51,126.7) .. controls (234.79,126.7) and (234.21,126.2) .. (234.21,125.6) -- cycle ;
\draw  [fill={rgb, 255:red, 0; green, 0; blue, 0 }  ,fill opacity=1 ] (245.87,125.6) .. controls (245.87,124.99) and (246.45,124.5) .. (247.17,124.5) .. controls (247.89,124.5) and (248.47,124.99) .. (248.47,125.6) .. controls (248.47,126.2) and (247.89,126.7) .. (247.17,126.7) .. controls (246.45,126.7) and (245.87,126.2) .. (245.87,125.6) -- cycle ;
\draw [color={rgb, 255:red, 255; green, 0; blue, 0 }  ,draw opacity=1 ]   (82.7,126.7) -- (111.95,160.07) ;
\draw [color={rgb, 255:red, 255; green, 0; blue, 0 }  ,draw opacity=1 ]   (94.41,126.7) -- (111.95,160.07) ;
\draw [color={rgb, 255:red, 255; green, 0; blue, 0 }  ,draw opacity=1 ]   (100.12,125.6) -- (111.95,160.07) ;
\draw [color={rgb, 255:red, 255; green, 0; blue, 0 }  ,draw opacity=1 ]   (117.84,126.7) -- (111.95,160.07) ;
\draw [color={rgb, 255:red, 255; green, 0; blue, 0 }  ,draw opacity=1 ]   (123.54,126.7) -- (111.95,160.07) ;
\draw [color={rgb, 255:red, 255; green, 0; blue, 0 }  ,draw opacity=1 ]   (141.26,126.7) -- (111.95,160.07) ;
\draw    (79.75,152.88) -- (79.56,142.71) ;
\draw [shift={(79.52,140.71)}, rotate = 88.93] [color={rgb, 255:red, 0; green, 0; blue, 0 }  ][line width=0.75]    (4.37,-1.32) .. controls (2.78,-0.56) and (1.32,-0.12) .. (0,0) .. controls (1.32,0.12) and (2.78,0.56) .. (4.37,1.32)   ;
\draw [color={rgb, 255:red, 255; green, 0; blue, 0 }  ,draw opacity=1 ]   (111.95,160.07) -- (112,176.38) ;
\draw    (150.5,152.88) -- (150.31,142.71) ;
\draw [shift={(150.27,140.71)}, rotate = 88.93] [color={rgb, 255:red, 0; green, 0; blue, 0 }  ][line width=0.75]    (4.37,-1.32) .. controls (2.78,-0.56) and (1.32,-0.12) .. (0,0) .. controls (1.32,0.12) and (2.78,0.56) .. (4.37,1.32)   ;
\draw    (220.75,152.88) -- (220.56,142.71) ;
\draw [shift={(220.52,140.71)}, rotate = 88.93] [color={rgb, 255:red, 0; green, 0; blue, 0 }  ][line width=0.75]    (4.37,-1.32) .. controls (2.78,-0.56) and (1.32,-0.12) .. (0,0) .. controls (1.32,0.12) and (2.78,0.56) .. (4.37,1.32)   ;
\draw    (291,152.88) -- (290.81,142.71) ;
\draw [shift={(290.77,140.71)}, rotate = 88.93] [color={rgb, 255:red, 0; green, 0; blue, 0 }  ][line width=0.75]    (4.37,-1.32) .. controls (2.78,-0.56) and (1.32,-0.12) .. (0,0) .. controls (1.32,0.12) and (2.78,0.56) .. (4.37,1.32)   ;
\draw  [fill={rgb, 255:red, 0; green, 0; blue, 0 }  ,fill opacity=1 ] (81.4,125.6) .. controls (81.4,124.99) and (81.98,124.5) .. (82.7,124.5) .. controls (83.42,124.5) and (84,124.99) .. (84,125.6) .. controls (84,126.2) and (83.42,126.7) .. (82.7,126.7) .. controls (81.98,126.7) and (81.4,126.2) .. (81.4,125.6) -- cycle ;
\draw  [fill={rgb, 255:red, 0; green, 0; blue, 0 }  ,fill opacity=1 ] (93.11,125.6) .. controls (93.11,124.99) and (93.7,124.5) .. (94.41,124.5) .. controls (95.13,124.5) and (95.72,124.99) .. (95.72,125.6) .. controls (95.72,126.2) and (95.13,126.7) .. (94.41,126.7) .. controls (93.7,126.7) and (93.11,126.2) .. (93.11,125.6) -- cycle ;
\draw  [fill={rgb, 255:red, 0; green, 0; blue, 0 }  ,fill opacity=1 ] (116.54,125.6) .. controls (116.54,124.99) and (117.12,124.5) .. (117.84,124.5) .. controls (118.56,124.5) and (119.14,124.99) .. (119.14,125.6) .. controls (119.14,126.2) and (118.56,126.7) .. (117.84,126.7) .. controls (117.12,126.7) and (116.54,126.2) .. (116.54,125.6) -- cycle ;
\draw  [fill={rgb, 255:red, 0; green, 0; blue, 0 }  ,fill opacity=1 ] (122.24,125.6) .. controls (122.24,124.99) and (122.82,124.5) .. (123.54,124.5) .. controls (124.26,124.5) and (124.84,124.99) .. (124.84,125.6) .. controls (124.84,126.2) and (124.26,126.7) .. (123.54,126.7) .. controls (122.82,126.7) and (122.24,126.2) .. (122.24,125.6) -- cycle ;
\draw  [fill={rgb, 255:red, 0; green, 0; blue, 0 }  ,fill opacity=1 ] (139.96,125.6) .. controls (139.96,124.99) and (140.55,124.5) .. (141.26,124.5) .. controls (141.98,124.5) and (142.57,124.99) .. (142.57,125.6) .. controls (142.57,126.2) and (141.98,126.7) .. (141.26,126.7) .. controls (140.55,126.7) and (139.96,126.2) .. (139.96,125.6) -- cycle ;
\draw  [fill={rgb, 255:red, 0; green, 0; blue, 0 }  ,fill opacity=1 ] (98.82,125.6) .. controls (98.82,124.99) and (99.4,124.5) .. (100.12,124.5) .. controls (100.84,124.5) and (101.42,124.99) .. (101.42,125.6) .. controls (101.42,126.2) and (100.84,126.7) .. (100.12,126.7) .. controls (99.4,126.7) and (98.82,126.2) .. (98.82,125.6) -- cycle ;
\draw  [color={rgb, 255:red, 74; green, 144; blue, 226 }  ,draw opacity=1 ] (201.25,125.17) .. controls (201.25,118.42) and (209.7,112.96) .. (220.13,112.96) .. controls (230.55,112.96) and (239,118.42) .. (239,125.17) .. controls (239,131.91) and (230.55,137.38) .. (220.13,137.38) .. controls (209.7,137.38) and (201.25,131.91) .. (201.25,125.17) -- cycle ;
\draw [color={rgb, 255:red, 74; green, 144; blue, 226 }  ,draw opacity=1 ]   (233.75,134) .. controls (245.25,142.88) and (254.25,158.13) .. (241.25,175.13) ;

\draw (99,177) node [anchor=north west][inner sep=0.75pt]   [align=center] {{\small Violation}};

\draw (222,177) node [anchor=north west][inner sep=0.75pt]   [align=left] {{\small No Violation}};

\draw (70,153) node [anchor=north west][inner sep=0.75pt]   [align=left] {{\small 06 May}};
\draw (71.5,159) node [anchor=north west][inner sep=0.75pt]   [align=left] {{\small 6 AM}};

\draw (140.34,153) node [anchor=north west][inner sep=0.75pt]   [align=left] {{\small 07 May}};
\draw (141.84,159) node [anchor=north west][inner sep=0.75pt]   [align=left] {{\small 6 AM}};

\draw (210.68,153) node [anchor=north west][inner sep=0.75pt]   [align=left] {{\small 08 May}};
\draw (212.18,159) node [anchor=north west][inner sep=0.75pt]   [align=left] {{\small 6 AM}};

\draw (281.02,153) node [anchor=north west][inner sep=0.75pt]   [align=left] {{\small 09 May}};
\draw (282.52,159) node [anchor=north west][inner sep=0.75pt]   [align=left] {{\small 6 AM}};

\draw (98,100) node [anchor=north west][inner sep=0.75pt]   [align=left] {Window 1};
\draw (168,100) node [anchor=north west][inner sep=0.75pt]   [align=left] {Window 2};
\draw (240,100) node [anchor=north west][inner sep=0.75pt]   [align=left] {Window 3};

\end{tikzpicture}
    \caption{Illustration of a time constraint with $m:t:s = 5:24\textrm{h}:\textrm{06 May 6 AM}$. In the first window we have a violation because there are $6$ occurrences of the option. Despite having $6$ close occurrences between the second and the third window, here we have no violations because they are not present within the same window.}
    \label{fig:timeconst}
\end{figure}

Each order $o$ has an associated set of constraints $C_o$, where each constraint $c \in C_o$ has a weight $w_c$. $C_o$ is the subset of constraints where the CNF mentioned above is satisfied by the combination of features of $o$. There are two types of constraints: \emph{window constraints} and \emph{time constraints}. Each window constraint can be encoded with a pair, also called \emph{rule}, of the form $m:n$. It means that only $m$ orders with that option (or set of options) can appear in a window of $n$ orders. This type of constraint is the most common in the literature. However, in our real-world scenario, we must also obey constraints based on time, e.g., to align with shifts or working days.
The time constraints can be encoded with the ternary $m:t:s$. Given the starting time $s$, consider a time window of duration $t$ starting from $s$ and ending at $s + t$. The next time window starts at $s + t$ and ends at $s + 2t$. We consider all the time windows $[s + \ell t, s + (\ell + 1) t)$ for all the integers $\ell$. The rule $m:t:s$ says that in any of these time windows, only $m$ orders can appear (Figure~\ref{fig:timeconst}).
Given an order $o$ at time $t'$ and the sequence of orders $S$ that already left the \bodybuffer, we compute the constraint violations of $o$ as follows.

\begin{enumerate}
    \item Initialize the violations of $o$ to $0$, i.e., $v_o \gets 0$.
    \item For each constraint $c$ in $C_o$:\\
    \phantom{M} Increase $v_o$ by $w_c$ if
    \setlength{\leftmarginii}{4em}
    \begin{enumerate}
        \item $c$ is a window constraint $m:n$ and the first $n-1$ elements of $S$ have at least $m$ orders matching constraint $c$, or
        \item $c$ is a time constraint and there are at least $m$ orders matching constraint $c$ that left the \bodybuffer{} in the time window $[s + \ell t, s + (\ell + 1) t)$, where $\ell$ is such that $t' \in [s + \ell t, s + (\ell + 1) t)$.
    \end{enumerate}
\end{enumerate}

\subsubsection{Built-to-Date}

Given the possibility of interruptions in the plant, it is not always guaranteed that the vehicles are always fully assembled on time for delivery. Therefore, we prioritize older orders in the sequence we produce. More formally, we aim to improve the so-called \emph{built-to-date} KPI, which measures the deviation of the actual build date of a car from the planned date, which could negatively affect customer satisfaction and logistics of supplies that are supposed to be delivered just in time.
Our interaction with the plant takes place in the very early stages, and we have no control over the dynamics in the Paint Shop or Final Assembly. However, we can reduce the spread factor of the cars planned for a specific date. Since cars that are ``close'' together in the sequence are more likely to be produced on the same day, we increase the chance of satisfying the \emph{built-to-date} KPI.

If possible, we also want to make the arrangement of cars more similar to the sequence in which they were planned, which we refer to as the build sequence. This increases the chance of meeting the supply requirements at the Assembly stage, as supplies are delivered just-in-time according to the build sequence.

\subsection{Algorithm}
\label{algorithm}

The key actions of our algorithm are triggered by three events for each car: lane enqueuing, lane dequeuing, and substitution (Figure~\ref{fig:enqdeqsub}).

\subsubsection{Setting}\label{setting}

The input and output of our algorithm is a stream of pairs consisting of a partially built car and an order. An order is a logical object: the request to build a specific car with specific features. A partially built car represents a physical object in which some of the features of the order are already realized, but some are not. After the Body Shop, only the type of the chassis is fixed, and after paint, the color of the car is also fixed. Early in the production, a partially built car can still be matched with many different orders, and as the production proceeds, the number of possible matches shrinks. We adapt the matching between partially built cars and orders dynamically in a process called \emph{substitution}. This is the core operation of what is called \emph{virtual resequencing}~\cite{resequencing_boysen_2012,SUN2024109990} in the literature.

\begin{figure}[htb!]
    \centering
    \includegraphics[width=0.8\textwidth]{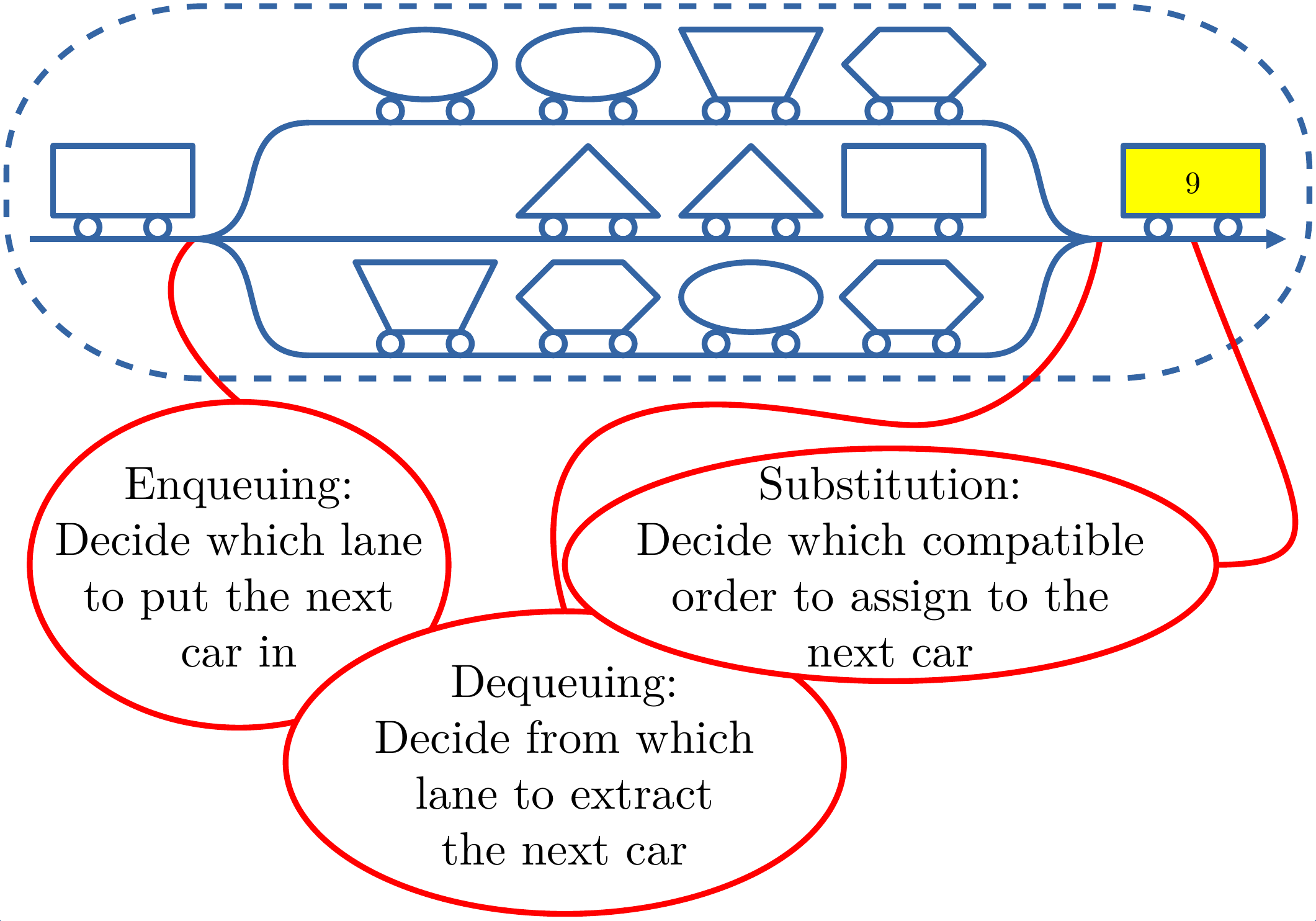}
    \caption{Key events that trigger the control of the buffer. During the substitution, we assign a compatible order to the car. In this case, the order has to match the rectangular body type. The order also encodes the color (yellow), the \blendnumber{} ($9$), and other features.}
    \label{fig:enqdeqsub}
  \end{figure}

Via a REST API\footnote{\url{https://gitlab.mpi-klsb.mpg.de/karrenba/modigpro}}, our algorithm controls the operation of the \bodybuffer{} and consists of three main components (Figure~\ref{fig:enqdeqsub}).

\begin{enumerate}[(i)]
    \item \textbf{Enqueuing:} Choose a lane for a car entering the \bodybuffer. 
    \item \textbf{Dequeuing:} Choose the next car to leave the \bodybuffer.
    \item \textbf{Substitution:} Reassign a different order to a car leaving the \bodybuffer.
\end{enumerate}

Our goal is to produce a sequence that balances multiple objectives. Specifically, we strive to achieve a favorable arrangement of colors, minimize constraint violations, and minimize delays in delivering cars to customers. We discuss these objectives in Section~\ref{subsec:kpi}.

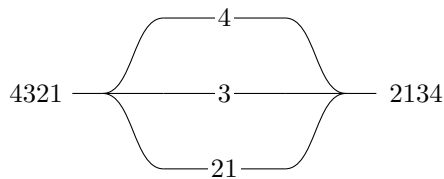
\begin{figure}[htb!]
    \centering
\begin{tikzpicture}

    \node[left=0.5cm] at (0.5,0) (4321) {$4 3 2 1$};
    \node[left=0.5cm] at (5.5,0) (2134) {$2 1 3 4$};

    \pgfmathsetmacro{\numlines}{3}
    \pgfmathsetmacro{\maxheight}{2}
    \pgfmathsetmacro{\xleft}{0} %
    \pgfmathsetmacro{\xright}{4} %
    \pgfmathsetmacro{\yverticalsegments}{0}
    \pgfmathsetmacro{\totlen}{\xright - \xleft} %

    \def\ratioSegment{1}
    \def\ratioCurve{2}
    \def\ratioLines{4}
    \def\totalRatio{10}

    \pgfmathsetmacro{\xLeftSegmentEnd}{\xleft + \ratioSegment * \totlen / \totalRatio}
    \pgfmathsetmacro{\xLeftCurveStart}{\xLeftSegmentEnd}
    \pgfmathsetmacro{\xParallelLinesStart}{\xleft + (\ratioSegment + \ratioCurve) * \totlen / \totalRatio}
    \pgfmathsetmacro{\xParallelLinesEnd}{\xleft + (\ratioSegment + \ratioCurve + \ratioLines) * \totlen / \totalRatio}
    \pgfmathsetmacro{\xRightCurveStart}{\xParallelLinesEnd}
    \pgfmathsetmacro{\xRightSegmentStart}{\xleft + (\ratioSegment + \ratioCurve + \ratioLines + \ratioCurve) * \totlen / \totalRatio}
    \pgfmathsetmacro{\xRightSegmentEnd}{\xright}

    \ifnum\numlines>1
        \pgfmathsetmacro{\linesep}{\maxheight/(\numlines-1)}
    \else
        \def\linesep{0}
    \fi

    \draw (\xleft,\yverticalsegments) -- (\xLeftSegmentEnd,\yverticalsegments) coordinate (left_vert);
    \draw (\xRightSegmentStart,\yverticalsegments) -- (\xRightSegmentEnd,\yverticalsegments) coordinate (right_vert);

        \pgfmathsetmacro{\currentshift}{\yverticalsegments - \maxheight/2}
        
        \draw (\xParallelLinesStart,{\currentshift}) -- (\xParallelLinesEnd,{\currentshift})
          node[pos=0.5, fill=white, inner sep=1pt] {$2 1$}
          coordinate (hline-1-right);
        
        \draw (\xLeftSegmentEnd,\yverticalsegments) .. controls (\xLeftCurveStart + 0.5*\ratioCurve * \totlen / \totalRatio,\yverticalsegments) and (\xParallelLinesStart - 0.5*\ratioCurve * \totlen / \totalRatio,{\currentshift}) .. (\xParallelLinesStart,{\currentshift});
        
        \draw (\xParallelLinesEnd,{\currentshift}) .. controls (\xParallelLinesEnd + 0.5*\ratioCurve * \totlen / \totalRatio,\currentshift) and (\xRightSegmentStart - 0.5*\ratioCurve * \totlen / \totalRatio,\yverticalsegments) .. (\xRightSegmentStart,\yverticalsegments);

        \pgfmathsetmacro{\currentshift}{\yverticalsegments - \maxheight/2 + \linesep}
        \draw (\xParallelLinesStart,{\currentshift}) -- (\xParallelLinesEnd,{\currentshift})
        node[pos=0.5, fill=white, inner sep=1pt] {$3$}
        coordinate (hline-2-right);
        
        \draw (\xLeftSegmentEnd,\yverticalsegments) .. controls (\xLeftCurveStart + 0.5*\ratioCurve * \totlen / \totalRatio,\yverticalsegments) and (\xParallelLinesStart - 0.5*\ratioCurve * \totlen / \totalRatio,{\currentshift}) .. (\xParallelLinesStart,{\currentshift});
        \draw (\xParallelLinesEnd,{\currentshift}) .. controls (\xParallelLinesEnd + 0.5*\ratioCurve * \totlen / \totalRatio,\currentshift) and (\xRightSegmentStart - 0.5*\ratioCurve * \totlen / \totalRatio,\yverticalsegments) .. (\xRightSegmentStart,\yverticalsegments);

        \pgfmathsetmacro{\currentshift}{\yverticalsegments - \maxheight/2 + 2 * \linesep}
        \draw (\xParallelLinesStart,{\currentshift}) -- (\xParallelLinesEnd,{\currentshift})
        node[pos=0.5, fill=white, inner sep=1pt] {$4$}
        coordinate (hline-2-right);
        \draw (\xLeftSegmentEnd,\yverticalsegments) .. controls (\xLeftCurveStart + 0.5*\ratioCurve * \totlen / \totalRatio,\yverticalsegments) and (\xParallelLinesStart - 0.5*\ratioCurve * \totlen / \totalRatio,{\currentshift}) .. (\xParallelLinesStart,{\currentshift});
        \draw (\xParallelLinesEnd,{\currentshift}) .. controls (\xParallelLinesEnd + 0.5*\ratioCurve * \totlen / \totalRatio,\currentshift) and (\xRightSegmentStart - 0.5*\ratioCurve * \totlen / \totalRatio,\yverticalsegments) .. (\xRightSegmentStart,\yverticalsegments);
\end{tikzpicture}
    \caption{It is impossible to sort a decreasing sequence of length $4$ with $3$ lanes.}
    \label{fig:decreasing}
\end{figure}

The capabilities of rearranging a sequence with a collection of parallel lanes are limited. In a companion paper~\cite{karrenbauer2023improving}, we have studied how much the sortedness can be improved by a collection of parallel lanes. Assume we are given a sequence of numbers and our goal is to rearrange the numbers in increasing order. Clearly, if the input is a decreasing sequence of length $n$, and we have fewer than $n$ lanes, we cannot sort perfectly, as at least two of the numbers have to go to the same lane (Figure~\ref{fig:decreasing}). We have studied two measures of sortedness: the length of the longest decreasing subsequence (LDS) and the number of runs (= increasing subsequences) composing a sequence. For the former measure, we have shown that for any sequence with LDS $L$, it is always possible to reduce it to an LDS of at most $L-k+1$ using $k$ lanes. For the latter measure, we have an optimal algorithm. The companion paper assumes lanes of unbounded capacity and, more importantly, does not take substitution into account. However, substitutions are extremely powerful, and our theoretical results therefore had little direct impact on the implemented algorithm. However, they guided our thinking, and we use the LDS-measure for evaluating sequences, see Section~\ref{lane enqueuing}.

We proceed by presenting the three modules of our algorithm. Given the dependencies of these three decisions, it is more natural to present them in reverse order: substitution, lane dequeuing, and lane enqueuing.

\subsubsection{Substitution} \label{pairass}
Recall that we want to focus on three different aspects:
\begin{enumerate}
    \item Constraints being satisfied
    \item Built-to-date and \blendnumbers{} in increasing order
    \item Large (assessed) average batch size
\end{enumerate}

When a \carin{} is required to be paired with some \vin{}, we first gather all the available \vins{} that are compatible (e.g., that match its body type). Then, we follow the steps outlined below, which iteratively shrink the set of compatible candidate \vins{} until only a single \vin{} is left.
\begin{enumerate}
    \item We consider only the \vins{} having the fewest constraint violations.
    \item If there are multiple \vins{} sharing the same minimum value,\footnote{This actually happens often because we can avoid violations most of the time, i.e., many \vins{} yield a violation of $0$.} we proceed by considering only the earliest due date.
    \item Should there still be more than one \vin{} left, we consider only those belonging to the last $k$ colors, where $k$ is a flexible parameter that we determined empirically to yield the best results on historic data\footnote{In Section~\ref{Determining k}, we explain how we computed the best value for this parameter.}. For example, say we have $5$ colors in total: red (R), green (G), blue (B), yellow (Y), and white (W). Consider the sequence of colors that have left the \bodybuffer{} R, R, G, B, R, Y, W, where the elements on the right have left earlier and the element on the left have left later. In this case, the last $k = 3$ colors are R, G, and B; therefore, we would only consider \vins{} in red, green, and blue. If there are no \vins{} in the last $k$ colors, we skip step $3$ and proceed with step $4$.
    \item Finally, among the remaining \vins{}, we select the one with the smallest \blendnumber{}, which is unique.
\end{enumerate}

As the results in Section~\ref{results} show, these steps tackle all our objectives simultaneously. The first one takes care of the constraint violations, the second one minimizes delays in the delivery, and the third one reduces the color changeovers, or equivalently, it increases the aABS. The reason behind this last improvement may, however, be less intuitive. The idea behind it is that this strategy generates sequences where few colors appear in large windows. This facilitates the job of the smaller buffer in the Paint Shop to distribute the cars over the two paint lines. Moreover, this approach has been found to be robust against the additional distortions after the \bodybuffer.

\subsubsection{Lane Dequeuing}
At this stage, we have to suggest the best \carin{} to leave the \bodybuffer{} next. We are given the list of \carins{} that are currently located at the head of their queues and are ready to leave, i.e., are not blocked by the operator or by the control software.

\begin{enumerate}
    \item Compute the potential best \vins{} for each of the given \carins{} according to the steps described in Section~\ref{pairass}.
    \item Among these virtually assigned \vins{}, select the best one, again according to the steps in Section~\ref{pairass}.
    \item Select the \carin{} that is virtually paired with the best \vin{}.
    \item Should there be multiple \carins{} compatible with the same best \vin{}, select the one that has spent the most time in the \bodybuffer.
\end{enumerate}

The rationale for the tie breaker in the last step is that corrosion increases with the time spent in the buffer due to humidity in the air.

\subsubsection{Lane Enqueuing}\label{lane enqueuing}
When a car enters the \bodybuffer, the control software has to send it into one of the available lanes, i.e., one that is not full or locked by the operator or the system. This decision does not have an immediate impact on the resulting output sequence; however, there are consequences in the long term because, usually, several other cars have to leave the \bodybuffer{} in front of the car entering. Therefore, we select among the given available lanes the one with the best prediction for the resulting sequence when all cars currently in the buffer are dequeued. To this end, we virtually push the \carin{} at the back of each lane, and each time, we virtually dequeue all the cars from the \bodybuffer. When we dequeue all the cars from the \bodybuffer{}, we generate a sequence of potential \carin-\vin{} pairs. Given such a sequence, we can evaluate the total constraint violations. But what happens if there are sequences with the same constraint violations? We also have to rely on other measures. While we can efficiently quantify the quality of the output sequence in terms of how ``well sorted'' the \blendnumbers{} are, it is much harder to predict the average batch size when the cars go in for painting. For the former, we consider the LDS. %
For the latter, we consider the average batch size of the output sequence as a proxy for the average batch size that will be achieved in the Paint Shop because it can be computed efficiently, whereas predicting the latter (i.e., computing the assessed average batch size) requires a time-consuming simulation that we cannot afford in real-time. It is, however, important to remark that there is a statistically significant correlation between the average batch size after the \bodybuffer{} and the one in the Paint Shop. Therefore, the impact that this proxy has on the final sequence is expected to be positive. %
With this in mind, we have the components to define a tie-breaker to differentiate sequences with the same constraint violations. We compute the ratio $\frac{LDS}{ABS}$, which is always well defined as $ABS \geq 1$. Low values of this ratio indicate that the sequence has low LDS and high ABS.
Once the best sequence is identified, we propose the corresponding lane as the target for the designated \carin.

In summary:
\begin{enumerate}
    \item For each available lane $i$:
    \begin{enumerate}
        \item Virtually enqueue the car entering the \bodybuffer{} in the lane $i$.
        \item Virtually dequque all the cars in the \bodybuffer{} and generate a sequence $S_i$ of \carin-\vin{} pairs.
    \end{enumerate}
    \item Select the sequences resulting in the fewest constraint violations.
    \item If there are multiple sequences sharing the same minimum value, select only the sequence $S_i$ with the lowest value of $\frac{LDS}{ABS}$.
    \item Enqueue the car in lane $i$.
\end{enumerate}

\subsubsection{Substitution Alternatives}
\label{subsalts}
To test our preliminary ideas and analyze the impact of our theoretical research, we have developed a partial simulator of the plant. It reenacts the events that happened in the \bodybuffer, constantly keeping track of the position of each car and the availability of each order. In the simulation, one has control only over the substitution, but not over enqueuing or dequeuing cars from the \bodybuffer. Therefore, our starting point was to gather useful insights on the impact of different substitution strategies.
Given the necessity to prioritize the minimization of the constraint violations, the structure of the substitution algorithm has always been the same, i.e., we first address the constraints, and then we try and optimize the color distribution. The latter required some deeper understanding on which sequences are more likely to result in long batches after being reshuffled in the Paint Shop buffer.

First, we tried to prioritize each color by its number of appearances, i.e., choose the first available color by popularity.  %
Then, we analyzed what happens if we focus only on the last $k$ colors that left the \bodybuffer, prioritizing them by ``last appearance'', i.e., the last color seen has highest priority, the second to last has second highest priority, and so on. All the other colors would have lower priority and would be ranked by popularity. %
Then, we considered a slight modification of this last alternative, by giving the same priority to all the colors that do not belong to the last $k$ colors.  %
Our final and most successful evolution of this strategy, which we have presented above in detail, was to give the same priority to the first $k$ colors, disregarding their popularity and their exact last appearance within the $k$ colors.

We noticed that priority ranking based on popularity is not flexible enough. After successfully producing large batches out of the most popular colors, the algorithm would eventually be forced to prioritize these colors even when they are not available over a prolonged period, resulting in windows of cars where a few popular colors would appear many times at the beginning, and many unpopular colors would appear few times in random order in the end. Regardless of the sorting algorithm used in the Paint Shop, eventually this yields a sequence with many short batches of colors.

Our latest approach, on the other hand, ensures that only a few colors are used in large windows. It was inspired by known caching mechanisms that prioritize latest events. Notice that, by doing so, we could produce sequences with a very low ABS, even close to $1$. But since there are large windows with few colors, it is easier for the legacy sorting algorithm to produce large batches. The choice of the value of $k$ is based on several analyses that we present in the next section.

\subsubsection{The Number of Considered Last Colors}\label{Determining k}
 In the third step of substitution, see Section~\ref{pairass}, we consider only the last $k$ colors seen; we use $k = 3$. We now discuss our rationale for this value. We took a real input sequence of 23000 cars and fed it to the simulation of the \bodybuffer{} mentioned in Section~\ref{subsalts} for values of $k$ from $0$ to $5$, where $0$ means that no substitution is applied to the sequence. We then fed the output of the simulation to the Paint Shop simulator (Section~\ref{subsec:abs}) to evaluate the quality of different sequences. For the analysis, we measured the ABS after substitution, the distribution of batch lengths, the distribution of different colors in $50$-car windows (i.e., how many colors there are in any window of $50$ consecutive cars), and the aABS; the aABS is the \emph{target of optimization.}
  We recall that the Paint Shop has internal buffers that are operated by legacy code. The buffer immediately before the paint lanes, called the Primer Buffer (see Figure~\ref{fig:paintshop_scheme}), has six parallel lanes. So, if there were only six or fewer different colors (and buffer lanes were sufficiently long), there would be no need for optimization. However, the cars manufactured in the plant can be ordered in about 20 different colors, the actual number varying slightly from year to year. The aABS is  influenced by the distribution of batch lengths and the color differentiation. We will argue in the sequel that the latter has more impact than the former. 
  Intuitively, it may seem that longer batches correlate with fewer color changes, and maximizing the ABS might therefore seem the same as maximizing the aABS. However, imagine that the sequence that comes out of the \bodybuffer{} simply alternates between two colors. Then ABS is just 1, but the Primer Buffer can easily keep both paint lanes busy without any color changes. Also, having one long batch and all other batches of size one results in a fairly long average batch size, but many color changes. 

\begin{table}[htbp!]
    \centering
    \begin{tabular}{cc}
     \includegraphics[width=0.4\textwidth]{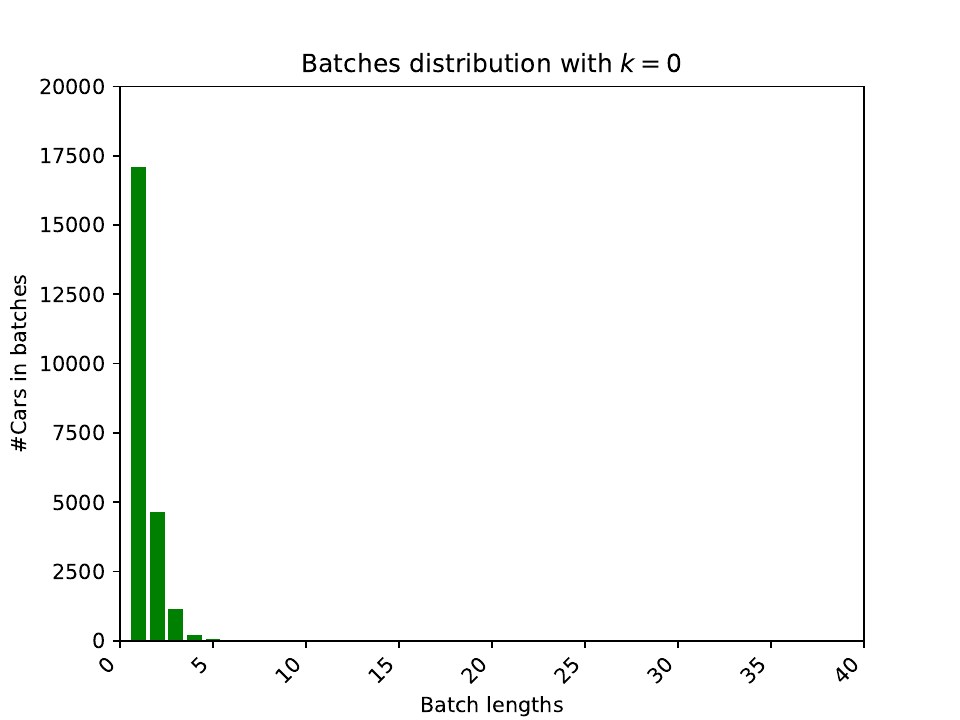}    & \includegraphics[width=0.4\textwidth]{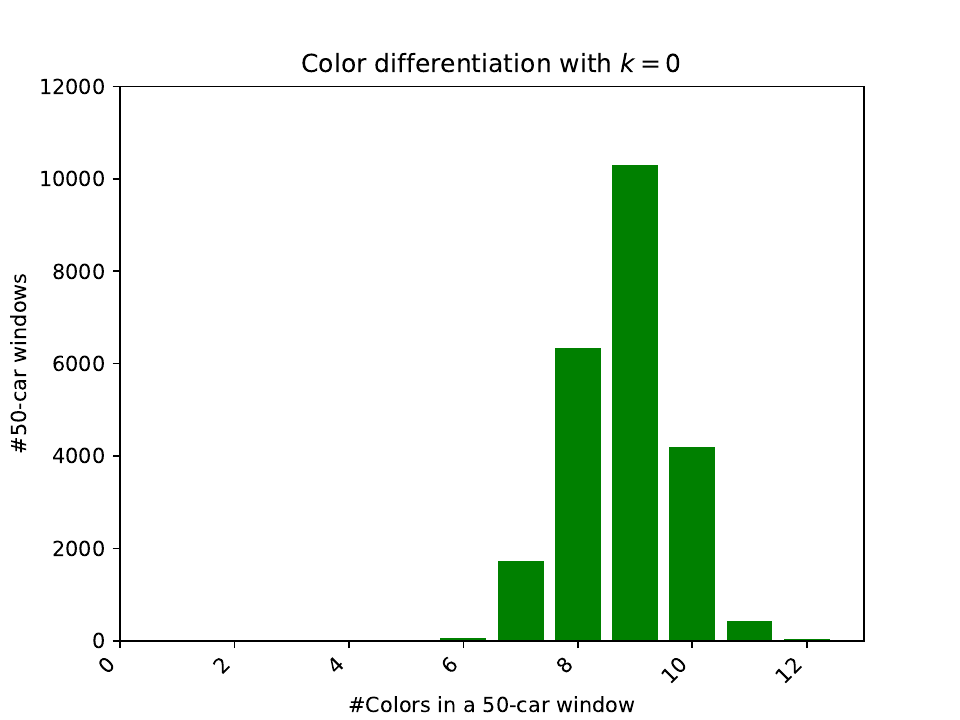} \\
     \includegraphics[width=0.4\textwidth]{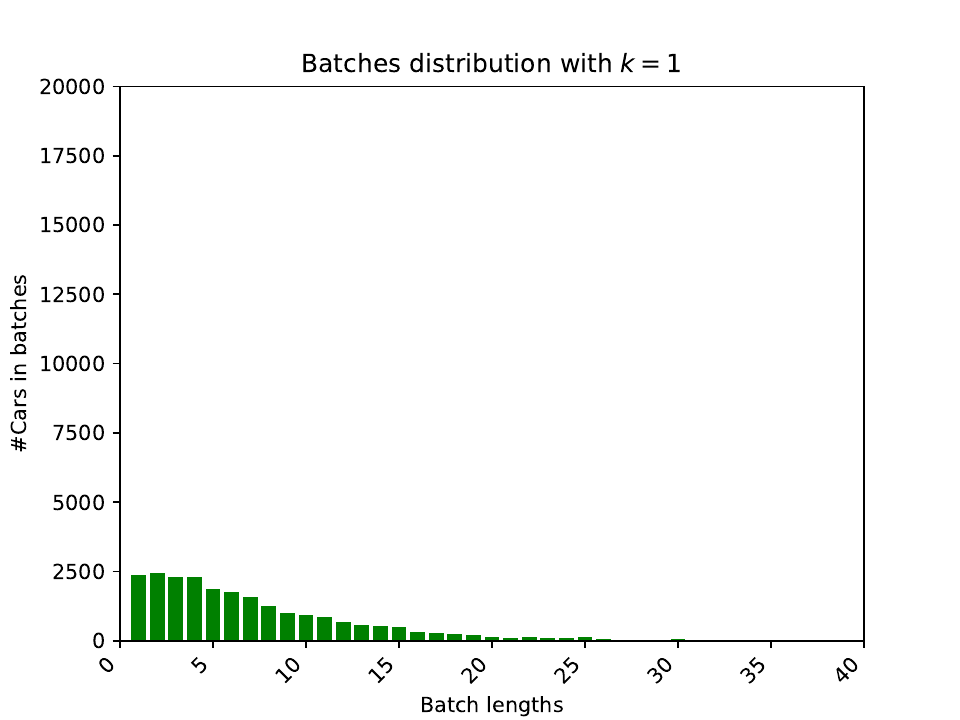}    & \includegraphics[width=0.4\textwidth]{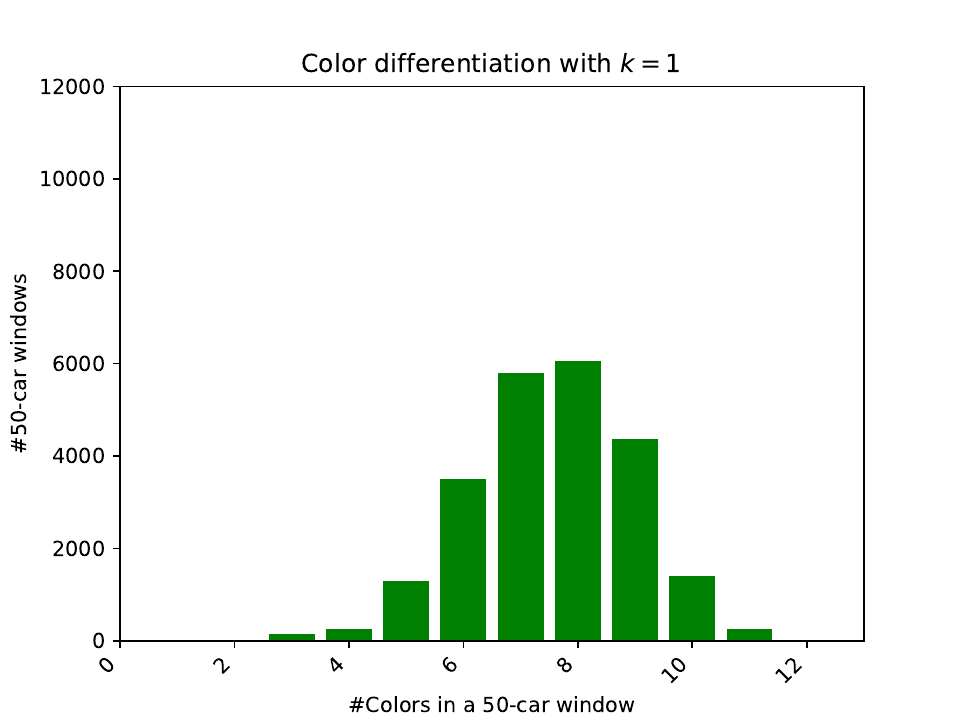} \\
     \includegraphics[width=0.4\textwidth]{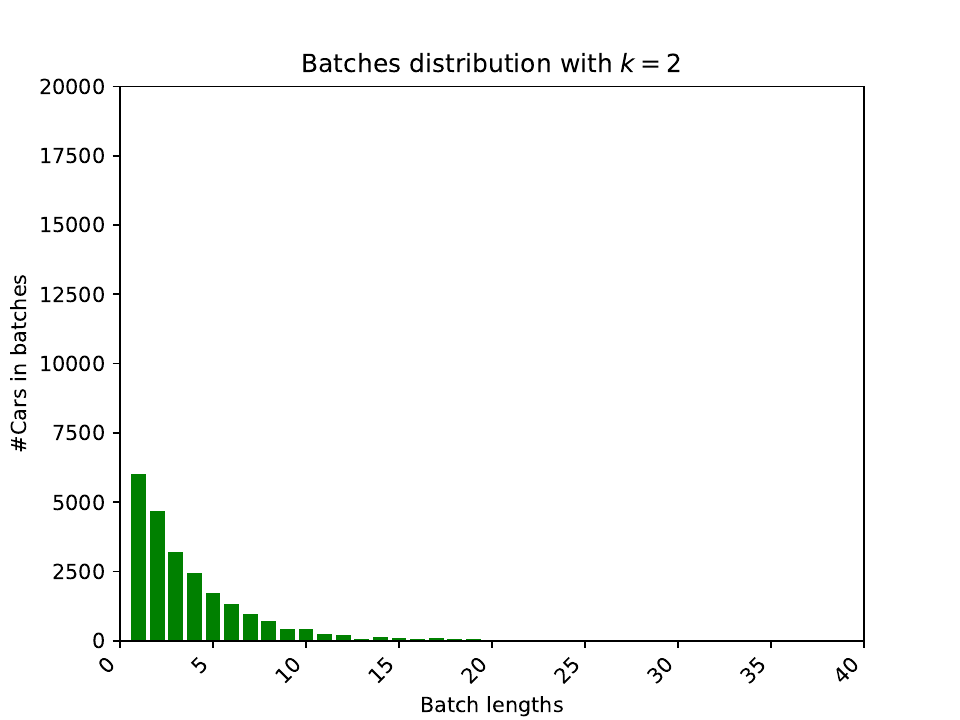}    & \includegraphics[width=0.4\textwidth]{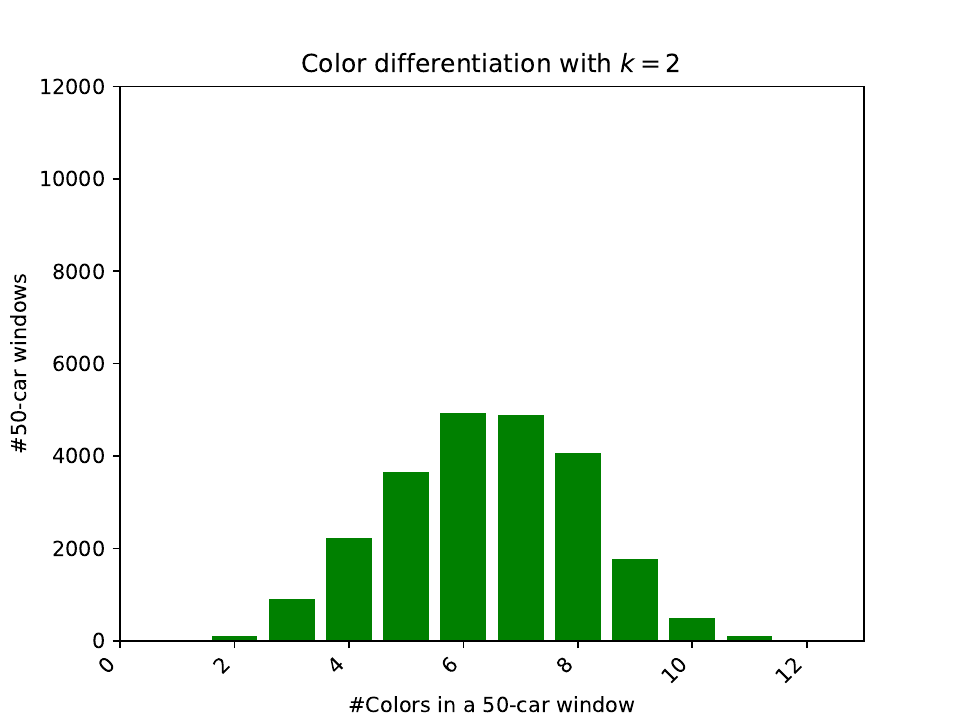}
    \end{tabular}
    \captionof{figure}{Distributions for $k=0, 1, 2$. On the left, the distributions showing the number of cars involved in each batch length. The less they are ``squeezed'' on the left, the better. On the right, the distribution of different colors in $50$-car windows. The more they are centered on low values, the better.}
    \label{fig:batchlengths0_2}
\end{table}

Figures~\ref{fig:batchlengths0_2} and~\ref{fig:batchlengths3_5} show the distributions of batch lengths and colors in $50$-car windows. They refer to the sequence before entering the Paint Shop. Recall that $k=0$ refers to the absence of substitutions. Once substitutions are used, the larger the $k$, the smaller the batch sizes. This is clearly visible in the diagrams on the left side of Figures~\ref{fig:batchlengths0_2} and~\ref{fig:batchlengths3_5}, respectively. The more we increase $k$, the more mass is moved to smaller batch sizes.
Does this not this suggest to use $k = 1$? No, we have already seen an example where batch size one leads to perfect usage of the paint lanes. The diagrams on the right side of the figure show the color differentiation, i.e, how many different colors occur in a 50-car window. It seems that favoring large average batch sizes leads to some long batches and also many short batches and, hence, many color changes.

\begin{table}[htbp]
    \centering
    \begin{tabular}{cc}
     \includegraphics[width=0.4\textwidth]{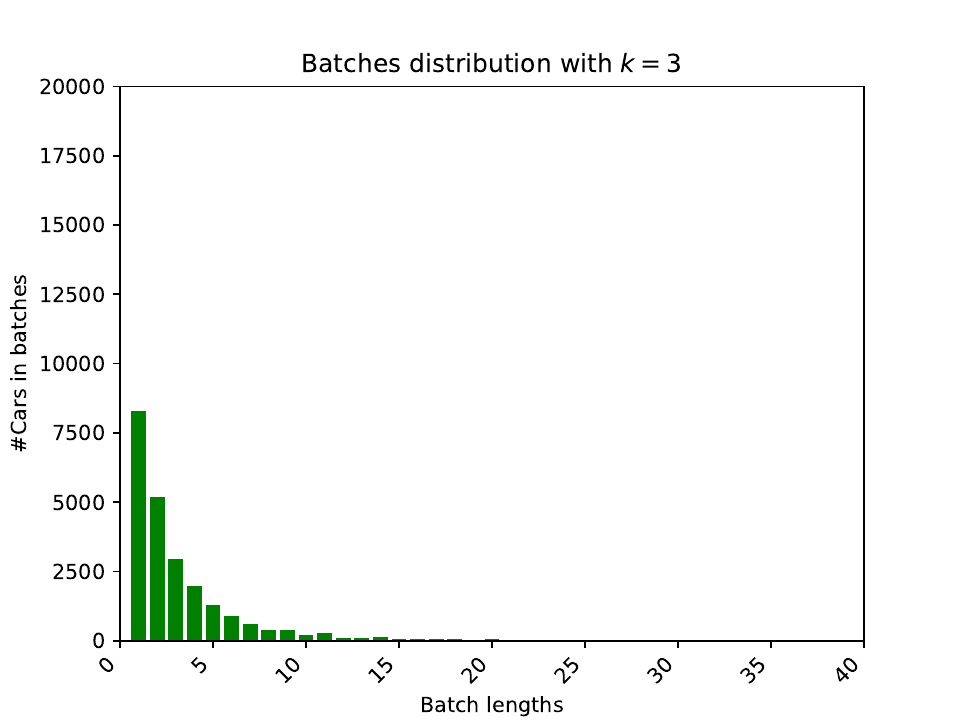}    & \includegraphics[width=0.4\textwidth]{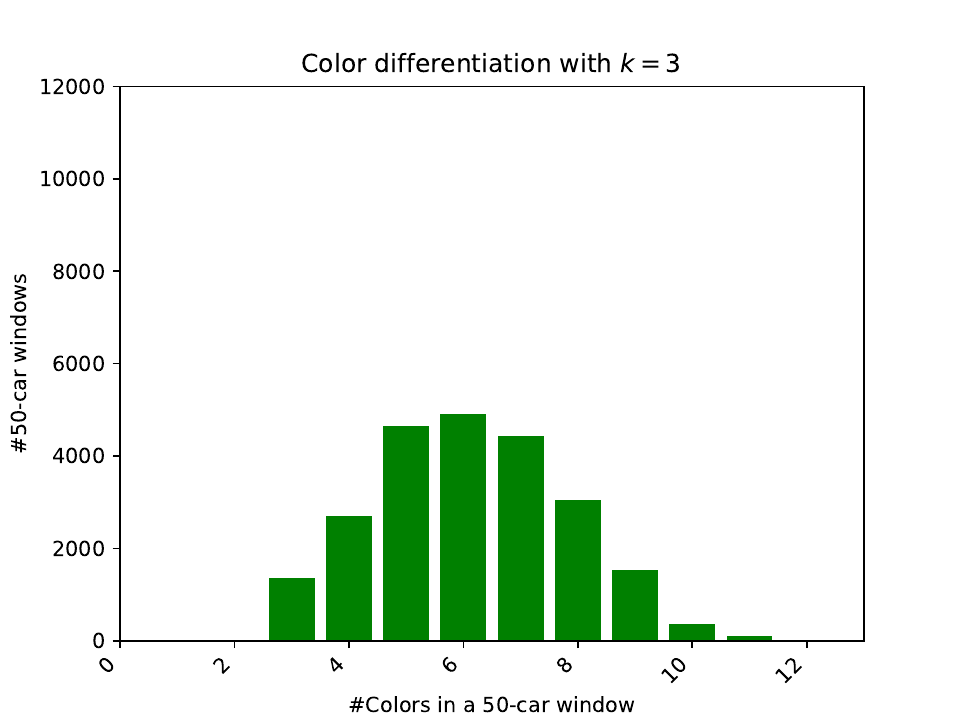} \\
     \includegraphics[width=0.4\textwidth]{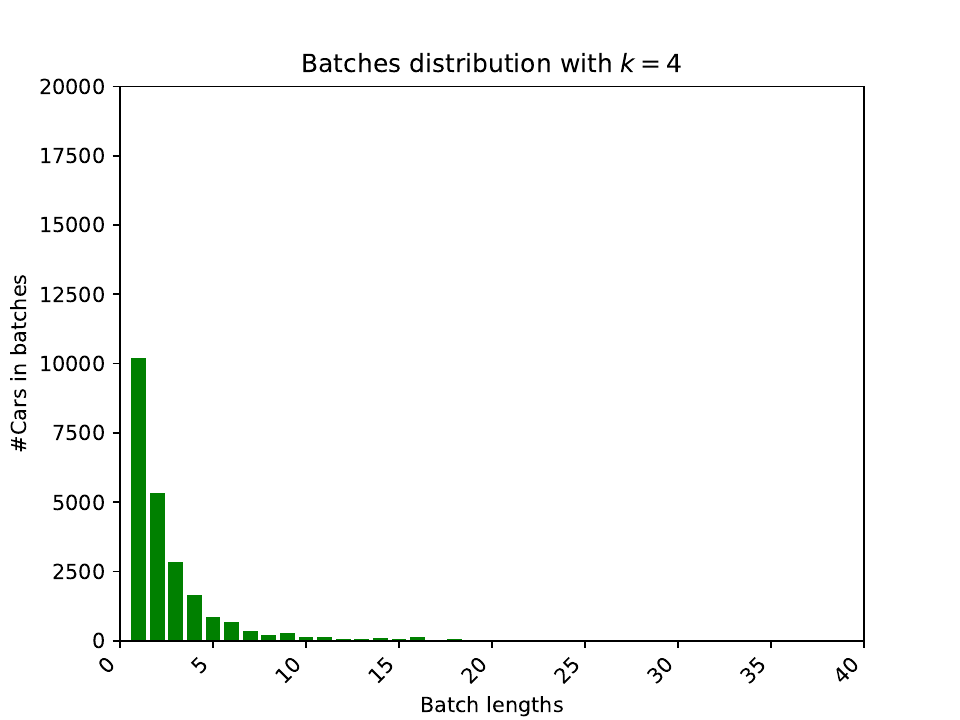}    & \includegraphics[width=0.4\textwidth]{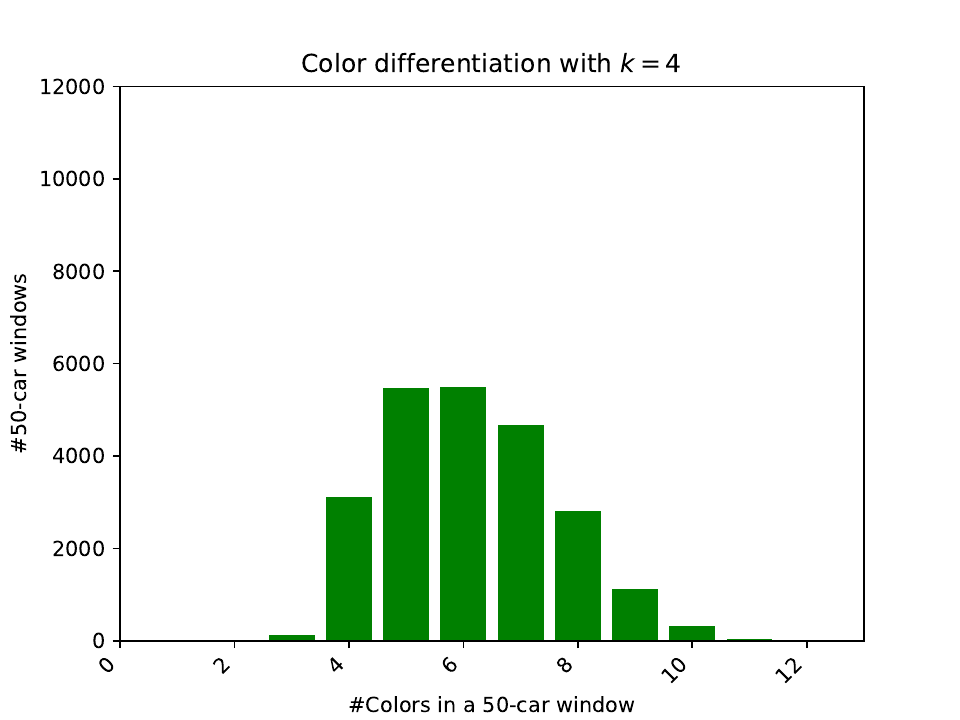} \\
     \includegraphics[width=0.4\textwidth]{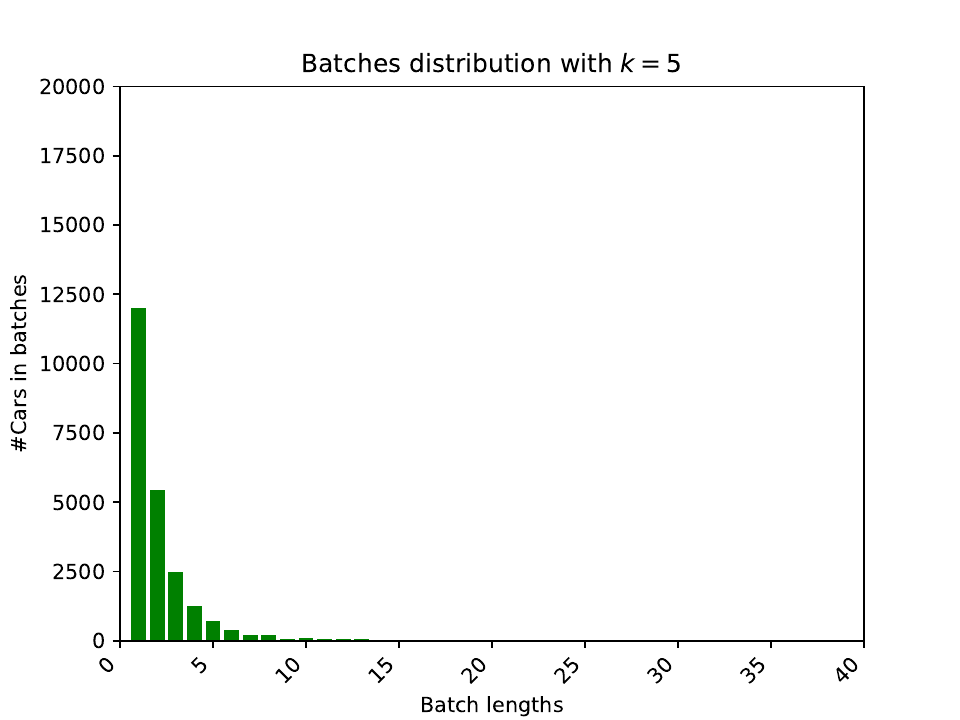}    & \includegraphics[width=0.4\textwidth]{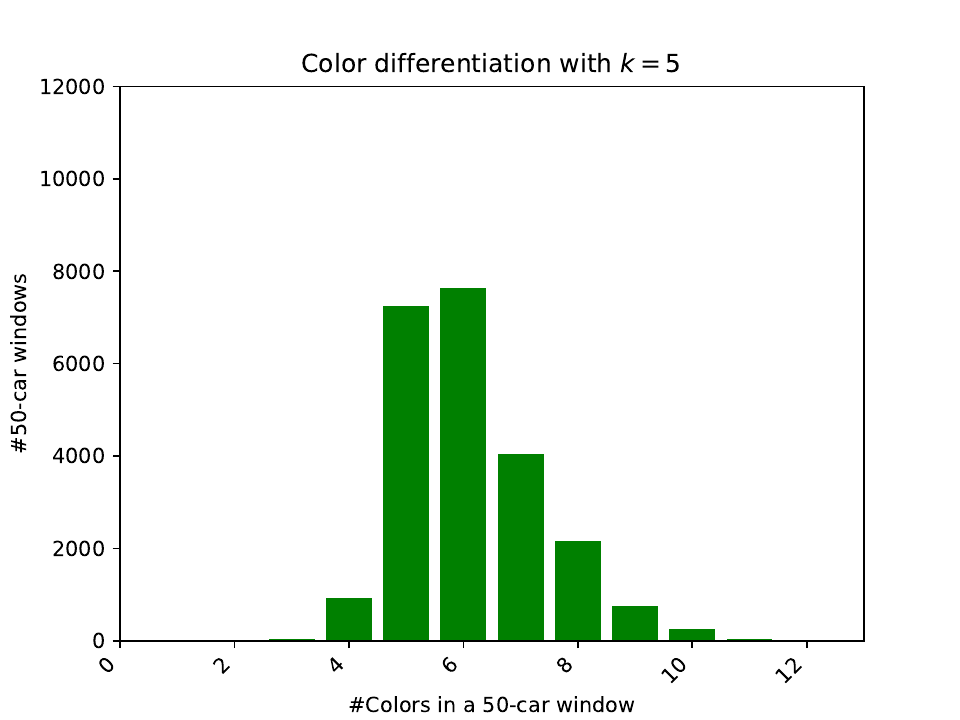}
    \end{tabular}
    \captionof{figure}{Distributions for $k=3, 4, 5$. On the left the distributions showing the number of cars involved in each batch length. Ignoring the Primer Buffer, short batches are to be avoided. On the right the distribution of different colors in $50$-car windows. The more they are concentrated on low values, the better.}
    \label{fig:batchlengths3_5}
\end{table}

\begin{figure}[htbp!]
    \centering
    \includegraphics[width=0.8\textwidth]{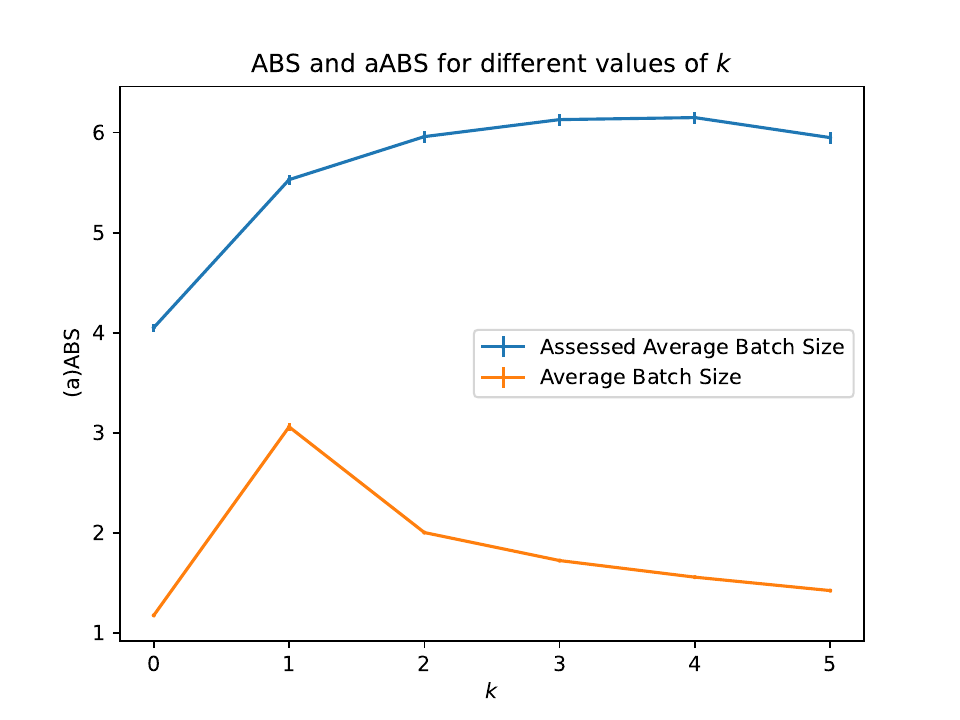}
    \caption{Values of Average Batch Size (ABS) and assessed Average Batch Size (aABS) for different values of $k$. The greater the aABS, the better. Standard deviations are also shown; they are tiny.} 
    \label{fig:abs_vs_k}
\end{figure}

As we said in the introduction, the actual target is the average batch size in the paint lanes. Figure~\ref{fig:abs_vs_k} shows the assessed Average Batch Size (aABS) for different values of $k$; the Average Batch Size (ABS) is also shown. The distribution of the color differentiation seems best for $k = 3$ and $k = 4$, as their respective aABS-distributions peak at lower values than for other values of $k$. On a statistical level, the two values are not really distinguishable. The tie-breaker, here, is the value of the ABS, which is considerably higher in the case $k=3$. Therefore, we decided to set the parameter $k$ to $3$. Figures~\ref{fig:batchlengths0_2} to~\ref{fig:abs_vs_k} indicate that color differentiation has a much larger impact on the performance of the Paint Shop than average batch size. 

\section{Results}\label{results}

After implementing the simulator mentioned in Section~\ref{subsalts} and making our design decisions based on historical data from the factory, we proceeded with the development of an API to integrate our work into the operation of the plant. Our industrial partner, in turn, developed a software named System Integrator that relays all relevant events to our API and dispatches the actions that we propose to the systems operating the plant.
Our API is written in Python using the FastAPI module.
The code is open-source and available at \url{https://gitlab.mpi-klsb.mpg.de/karrenba/modigpro}.

Our API has been operating in the plant since Q1 of 2023. During the first weeks, it had no concrete control over the plant's decisions, as it was only ``listening'' to the events, updating its internal status accordingly, and simulating the decisions it would make. %
It went live on 1 March 2023, followed by a hypercare phase of one month.

Every time the API receives an event or query, it records both the request and its response in a log file. We saved all the log files, thus, we have access to all the events that happened during these months. In order to analyze the performance of our system we collected the data of $20$ working days before and $20$ working days after out API went live in the plant, making a total of $8$ weeks of working days. To make the explanation easier, we will refer to the four weeks of evaluation before it became operational as period $P_{old}$, and the four weeks of evaluation after it became operational as period $P_{new}$. The total amount of events in the usable data that we collected amounts to $52973$. Since there are roughly two events per car (i.e., info while entering the \bodybuffer{} and info while leaving the \bodybuffer), the total number of cars produced in these two periods is more than $25000$. Despite having access to more data, it is crucial to acknowledge the limitations of our investigation's timeframe (8 weeks). Analyzing longer-term production trends is challenging because factory operations are dynamic. There are often ongoing adjustments (e.g., facility upgrades, shift changes) that can significantly impact results. Still, we observed stability within these 8 weeks, providing a clear basis for comparison between the two time periods.

At the end of the testing period, we analyzed the data to check if there were significant constraint violations and changes in the color distribution for the Paint Shop, in the \blendnumber{} distribution, and in the built-to-date distribution.

Since constraint violation is the criterion with highest priority, thus, considered before all other criteria, no severe constraint violations were observed. For this, we relied on the qualitative informal feedback by the operators in the plant because a quantitative evaluation of this aspect was not possible because we had no access to corresponding data. We have, however, reason to believe that the performance with respect to the constraints has remained stable.

The situation is different with respect to the paint colors and \blendnumbers{}. Though we have the data, it is not trivial to compare the color distribution of the sequence entering and leaving the \bodybuffer{} and assess any improvement of the quality. If our objective was to maximize the ABS right when leaving the \bodybuffer{}, it would suffice to compute this value on the sequence entering and leaving the \bodybuffer{} and observe how it increases. In our case, we are interested in reducing the ``complexity'' of the color distribution, where a complex distribution leads to a low aABS, while a simple distribution leads to a high aABS. We use the aABS as a measure of complexity of a sequence, and we can apply this measure even to the sequence entering the \bodybuffer. Although it has no practical meaning to compute the aABS on a sequence entering the \bodybuffer, it gives us insights on how adversarial the sequence is to the buffer entering the Paint Shop. This measure allows us to have a comparison of the quality of the sequences entering and leaving the \bodybuffer.

\subsection{Color Distribution}
\label{sec:coldist}
\begin{figure}[htb!]
  \centering
  \includegraphics[width=0.8\textwidth]{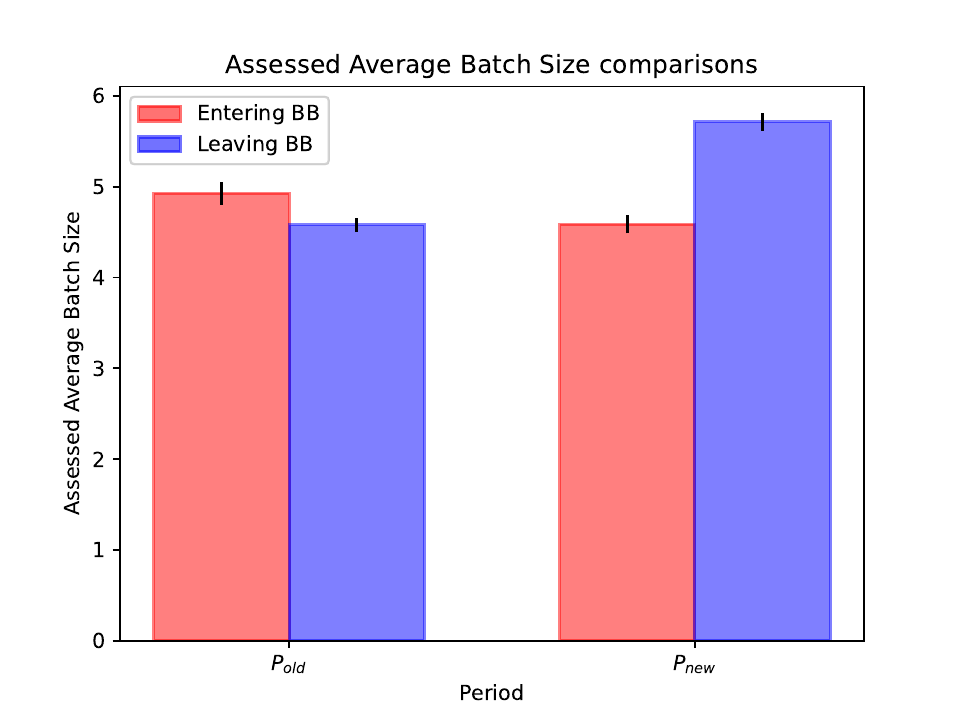}
  \caption{Assessed Average Batch Size (aABS) of the sequences entering and leaving the \bodybuffer{}. The higher the bin, the better. It is clear how during $P_{new}$ there is an evident improvement on the quality of the sequence with respect to the color distribution, whereas during $P_{old}$ the quality gets slightly worse.}
  \label{fig:abscomps}
\end{figure}

In Table~\ref{avgabs} and Figure~\ref{fig:abscomps} we present the mean values of the daily aABSs across the two periods $P_{old}$ and $P_{new}$., which are the periods before the API became active and after, respectively. Recall that the use of the aABS for the sequence entering the \bodybuffer{} serves only as a proxy for the complexity of the sequence.

\begin{table}[htb!]
\centering
\begin{tabular}{|c|ccc|ccc|}
\hline
\multirow{2}{*}{} & \multicolumn{3}{c|}{Entering \bodybuffer{}}                                        & \multicolumn{3}{c|}{Leaving \bodybuffer{}}                                         \\ \cline{2-7} 
                  & \multicolumn{1}{c|}{mean} & \multicolumn{1}{c|}{std}  & std error of the mean & \multicolumn{1}{c|}{mean} & \multicolumn{1}{c|}{std}  & std error of the mean \\ \hline
$P_{old}$             & \multicolumn{1}{c|}{4.93}    & \multicolumn{1}{c|}{0.57} & 0.13               & \multicolumn{1}{c|}{4.58}    & \multicolumn{1}{c|}{0.36} & 0.08               \\ \hline
$P_{new}$             & \multicolumn{1}{c|}{4.58}    & \multicolumn{1}{c|}{0.43} & 0.10               & \multicolumn{1}{c|}{5.71}    & \multicolumn{1}{c|}{0.65} & 0.15               \\ \hline
\end{tabular}
\caption{Mean values of assessed Average Batch Size (aABS) in the period before the API became active ($P_{old}$) and after ($P_{new}$). The higher the mean values, the better.}
\label{avgabs}
\end{table}

The fluctuating raw data makes it difficult to discern a trend between $P_{old}$ and $P_{new}$; however, this trend is readily apparent in Figure~\ref{fig:abscomps}, which visualizes mean values.
We can see that the sequence of cars entering the \bodybuffer{} (red bins), on average, preserves its complexity. In fact, it appears that during the second period the sequence of cars entering the \bodybuffer{} was slightly harder to handle, as the mean aABS is lower. During $P_{old}$, the complexity of the sequence leaving the \bodybuffer{} (blue bins) does not change significantly. This does not hold in $P_{new}$, as the mean aABS seems to increase substantially. More specifically, from this data, the improvement of the aABS of the sequences leaving the \bodybuffer{} between $P_{old}$ and $P_{new}$ is $24.7\%$.
This is just a semi-quantitative value because, as already discussed, it does not reflect the data in the plant, and we used approximated tools like our simulator. However, the plant operators reported an improvement of about $30\%$, which is not only in line with, but is even better than, our predictions.
\medskip\newline
To further emphasize this improvement, a $t$-test on the values of $P_{old}$ shows that the values of aABS decrease with a $p$-value of $0.0018$, considerably lower than $0.05$, which makes the claim statistically significant. For $P_{new}$ we first show a correlation between the values entering the \bodybuffer{} and those leaving it, and then we execute an orthogonal regression to quantify the improvement. The correlation is $0.47$ with $p$-value $0.035$. The low $p$-value guarantees a considerable correlation between the two sets of data and justifies the linear regression. The orthogonal regression results in a slope of $2.4 \pm 0.8$ and an intercept of $-5 \pm 4$, with a $p$-value of $0.049$. The value of the slope is remarkably high and demonstrates an impressive improvement on this side. \medskip\newline
We can now execute a similar analysis on the color changeovers. In Table~\ref{totalchanges}, we present the total number of cars analyzed and the total number of color changeovers.
\begin{table}[htbp!]
\centering
\resizebox{\textwidth}{!}{%
\begin{tabular}{|c|ccc|ccc|}
\hline
      & \multicolumn{3}{c|}{Entering \bodybuffer{}}                                             & \multicolumn{3}{c|}{Leaving \bodybuffer{}}                                              \\ \hline
      & \multicolumn{1}{c|}{Total cars} & \multicolumn{1}{c|}{Total color changeover} & CPC   & \multicolumn{1}{c|}{Total cars} & \multicolumn{1}{c|}{Total color changeover} & CPC   \\ \hline
$P_{old}$ & \multicolumn{1}{c|}{13668}      & \multicolumn{1}{c|}{2787}                & 0.204 & \multicolumn{1}{c|}{13610}      & \multicolumn{1}{c|}{2968}                & 0.218 \\ \hline
$P_{new}$ & \multicolumn{1}{c|}{11233}      & \multicolumn{1}{c|}{2451}                & 0.218 & \multicolumn{1}{c|}{11295}      & \multicolumn{1}{c|}{1982}                & 0.175 \\ \hline
\end{tabular}%
}
\caption{Total color changeover during $P_{old}$ and $P_{new}$. ``CPC'' stands for ``Changeover Per Car'', obtained by computing $\frac{\text{Total color changeover}}{\text{Total cars}}$. The lower this value, the better. The reason why the number of cars entering and leaving the \bodybuffer{} is not the same is because within the same day not all the cars that leave the \bodybuffer{} enter it the same day, and not all the cars that enter the \bodybuffer{} leave it the same day.}
\label{totalchanges}
\end{table}
We can see how the changeovers per car are lower with the new setup. For every $1000$ cars, the number of color changeovers is about $40$ changes lower than before. The table data indicates a $19.7\%$ improvement in the CPC leaving the \bodybuffer{}, comparing $P_{old}$ and $P_{new}$. Alternatively, we can calculate the CPC improvement using the plant's reported aABS improvement, observed to be $30\%$. The corresponding CPC improvement can be calculated as $1 - \frac{1}{0.3 + 1} = 23\%$. This provides again a comparison between the measured data and the plant's operational observations.

\subsection{Sequence Numbers Distribution and Built-to-Date}  %
While we can analyze the ABS and the number of color changeovers to evaluate the color distribution, it is harder to perform an evaluation of the \blendnumber{} distribution.
In order to quantify the ``sortedness'' of the sequences, we rely on three measures: LDS, expected value of the length of the decreasing subsequences, and median of the length of the decreasing subsequences. We already discussed the meaning of the LDS, which, from a practical point of view, tells us the number of lanes needed for a buffer to sort a sequence perfectly. Now, we explain more in detail the meaning of the other two measures. Within a single sequence there can be many decreasing subsequences. Assign to each element a longest decreasing subsequence ending with that element. For example, consider the sequence $4$, $1$, $3$, $5$, $2$. The longest decreasing subsequence that ends with $3$ is $4$, $3$, of length $2$, while the longest decreasing subsequece that ends with $2$ is $4$, $3$, $2$, of length $3$. The expected value of the length of the decreasing subsequences tells us ``on average'' what the length is of the longest decreasing subsequence ending with some element of the sequence. On the practical side, it tells us the number of lanes required for a \bodybuffer{} to perfectly sort ``most of'' the sequence. A similar explanation applies to the median. Given the distribution of the lengths of the subsequences ending with each element, it tells us at which value half of the elements have a corresponding decreasing subsequence of lower value and the other half of higher value.

How exactly do we analyze the behavior of these measures? When the sequence of cars enters the \bodybuffer{}, it is still in its early stages of the production process. %
Despite the limitations of the structure of the \bodybuffer{}, we show in Section~\ref{sec:coldist} that our work makes it possible to obtain good results in terms of colors. Improving the \blendnumbers{} arrangement represents a much more challenging goal to focus on, as it is often in conflict with our other goals. In fact, making decisions that satisfy critical criteria often backfires on the quality of \blendnumbers, which most of the times worsens in the resequencing process. To compare the change of the performance between different periods, we decided to report the ``average worsening'' of the sequence, i.e., by how much each measure gets worse when the sequence goes through the \bodybuffer{}. For all the measures, the worsening factor is computed by $\frac{\text{Value Leaving BB}}{\text{Value Entering BB}}$.

In Table~\ref{worseblend} and Figure~\ref{fig:wfcomps}, we report all the worsening factors, together with their standard deviation and the standard erorr of the mean.
In the Paint Shop, there are two lanes where the cars are painted in parallel. With ``Paintlane 1'' and ``Paintlane 2'' we refer to these to lanes (see Figure~\ref{fig:paintshop_scheme}).

\begin{table}[htb!]
\centering
\resizebox{\textwidth}{!}{%
\begin{tabular}{|cc|ccc|ccc|ccc|}
\hline
\multicolumn{2}{|c|}{\multirow{2}{*}{}}                    & \multicolumn{3}{c|}{LDS}                                              & \multicolumn{3}{c|}{Expected Value}                                        & \multicolumn{3}{c|}{Median}                                           \\ \cline{3-11} 
\multicolumn{2}{|c|}{}                                     & \multicolumn{1}{c|}{mean} & \multicolumn{1}{c|}{std}  & std of avg & \multicolumn{1}{c|}{mean} & \multicolumn{1}{c|}{std}  & std of avg & \multicolumn{1}{c|}{mean} & \multicolumn{1}{c|}{std}  & std of avg \\ \hline
\multicolumn{1}{|c|}{\multirow{2}{*}{Input}}       & $P_{old}$ & \multicolumn{1}{c|}{1.07}    & \multicolumn{1}{c|}{0.14} & 0.03       & \multicolumn{1}{c|}{1.27}    & \multicolumn{1}{c|}{0.11} & 0.02       & \multicolumn{1}{c|}{1.39}    & \multicolumn{1}{c|}{0.24} & 0.05       \\ \cline{2-11} 
\multicolumn{1}{|c|}{}                             & $P_{new}$ & \multicolumn{1}{c|}{0.92}   & \multicolumn{1}{c|}{0.12} & 0.03       & \multicolumn{1}{c|}{1.11}    & \multicolumn{1}{c|}{0.13} & 0.03       & \multicolumn{1}{c|}{1.24}    & \multicolumn{1}{c|}{0.18} & 0.04       \\ \hline
\multicolumn{1}{|c|}{\multirow{2}{*}{Paintlane 1}} & $P_{old}$ & \multicolumn{1}{c|}{1.12}    & \multicolumn{1}{c|}{0.20} & 0.05       & \multicolumn{1}{c|}{1.17}    & \multicolumn{1}{c|}{0.13} & 0.03       & \multicolumn{1}{c|}{1.39}    & \multicolumn{1}{c|}{0.24} & 0.05       \\ \cline{2-11} 
\multicolumn{1}{|c|}{}                             & $P_{new}$ & \multicolumn{1}{c|}{1.01}    & \multicolumn{1}{c|}{0.13} & 0.03       & \multicolumn{1}{c|}{1.10}    & \multicolumn{1}{c|}{0.11} & 0.02       & \multicolumn{1}{c|}{1.24}    & \multicolumn{1}{c|}{0.18} & 0.04       \\ \hline
\multicolumn{1}{|c|}{\multirow{2}{*}{Paintlane 2}} & $P_{old}$ & \multicolumn{1}{c|}{1.19}    & \multicolumn{1}{c|}{0.17} & 0.04       & \multicolumn{1}{c|}{1.26}    & \multicolumn{1}{c|}{0.12} & 0.03       & \multicolumn{1}{c|}{1.39}    & \multicolumn{1}{c|}{0.24} & 0.05       \\ \cline{2-11} 
\multicolumn{1}{|c|}{}                             & $P_{new}$ & \multicolumn{1}{c|}{1.00}    & \multicolumn{1}{c|}{0.13} & 0.03       & \multicolumn{1}{c|}{1.08}    & \multicolumn{1}{c|}{0.13} & 0.03       & \multicolumn{1}{c|}{1.24}    & \multicolumn{1}{c|}{0.18} & 0.04       \\ \hline
\end{tabular}
}
\caption{Mean worsening factor for LDS in comparison to Expected Value and Median of the length of the decreasing subsequences. The lower the mean value, the better.}
\label{worseblend}
\end{table}

\begin{table}[htb!]
\centering
\begin{tabular}{cc}
     \includegraphics[width=0.5\textwidth]{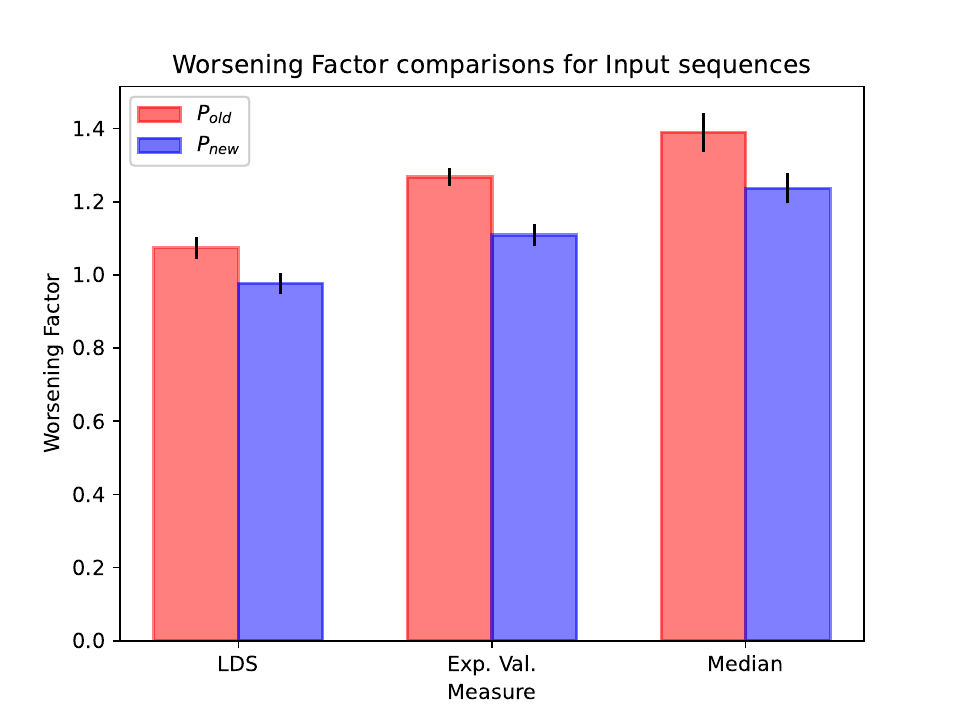} &  
     \includegraphics[width=0.5\textwidth]{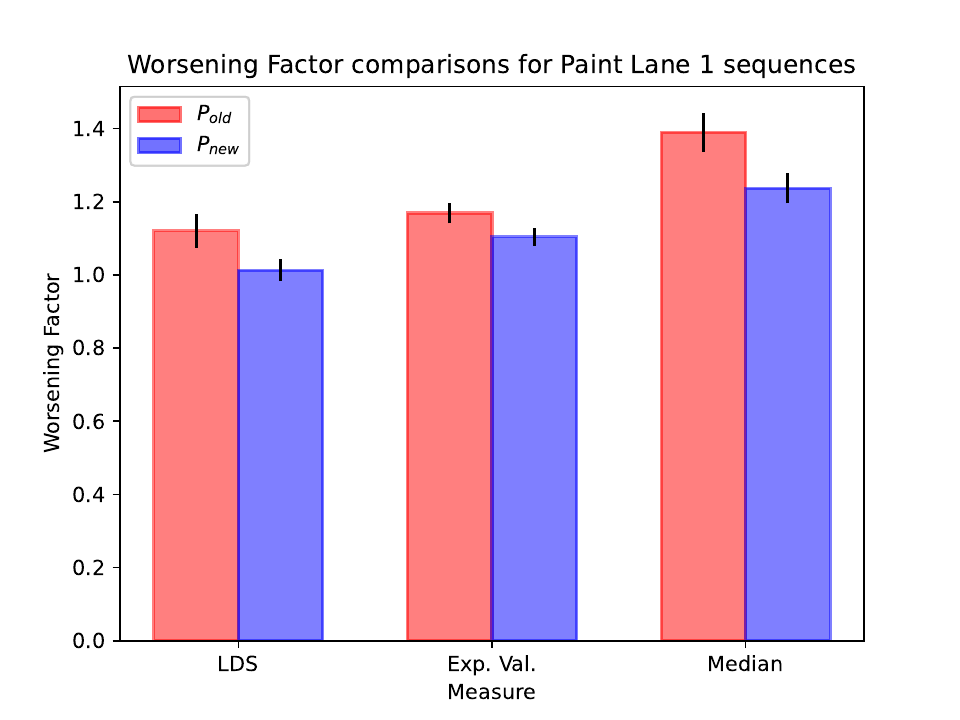} \\
     \multicolumn{2}{c}{\includegraphics[width=0.5\textwidth]{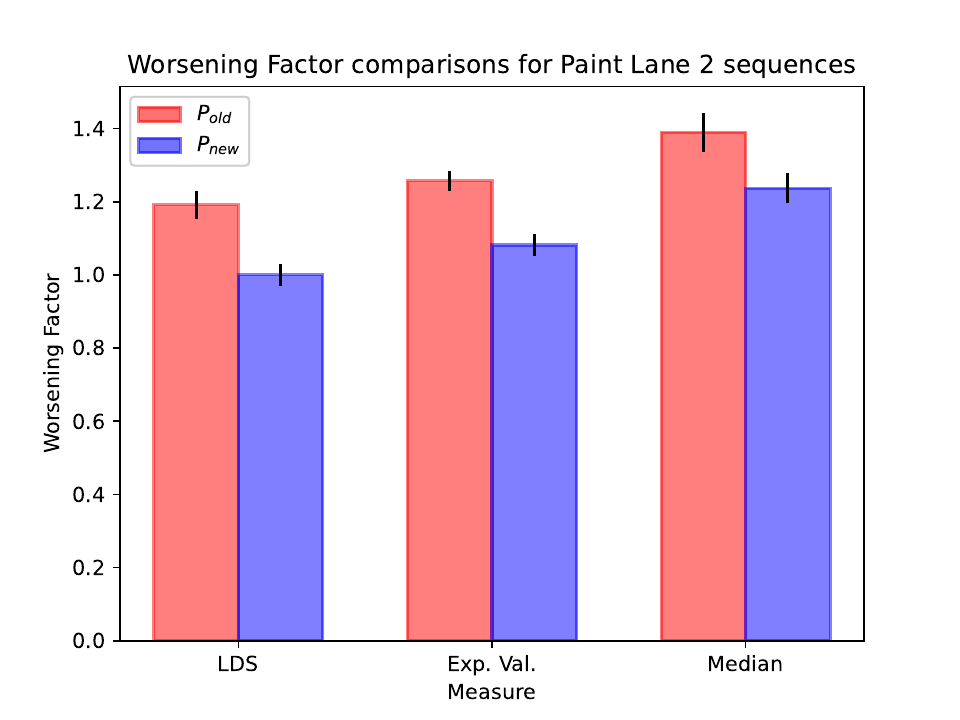}}
\end{tabular}
\captionof{figure}{Worsening factor comparisons. The lower the bin, the better. For any measure, it holds that the performance during $P_{new}$ (blue) is much better than during $P_{old}$ (red). The values of LDS for $P_{new}$ are compatible with $1$, indicating that the sequence's LDS was maintained when passing through the \bodybuffer{}.}
\label{fig:wfcomps}
\end{table}

By comparing $P_{old}$ and $P_{new}$, it is clear that in every case, on average, these measures undergo a much more reduced increment during $P_{new}$ with respect to $P_{old}$.
One can see that the values of the Median are identical within the same period. This unlikely event can happen because the values of the median are integers, and with a limited set of points, the overall results may end up being the same. \medskip\newline
What about the built-to-date? We cannot know if, at the end of the production line, the cars were built on the desired date, as we only operate at the very early stages. Nevertheless, we can still analyze how the cars planned for a specific date are distributed. In other words, we want the cars planned for a specific date to be ``close together'' in the production line so that the number of delayed cars is minimized. In order to quantify how close these cars are to each other, we compute the standard deviation of their indices in the sequence. We refer to this value as their \emph{index width}; for example, if a group of $100$ cars planned to be ready for a certain day is spread over a window of $1000$ cars, we expect a large \emph{index width}, while if they are spread over a window of $150$ cars, we expect a small \emph{index width}.

\begin{table}[htbp]
    \centering
    \begin{tabular}{cc}
     \includegraphics[width=0.5\textwidth]{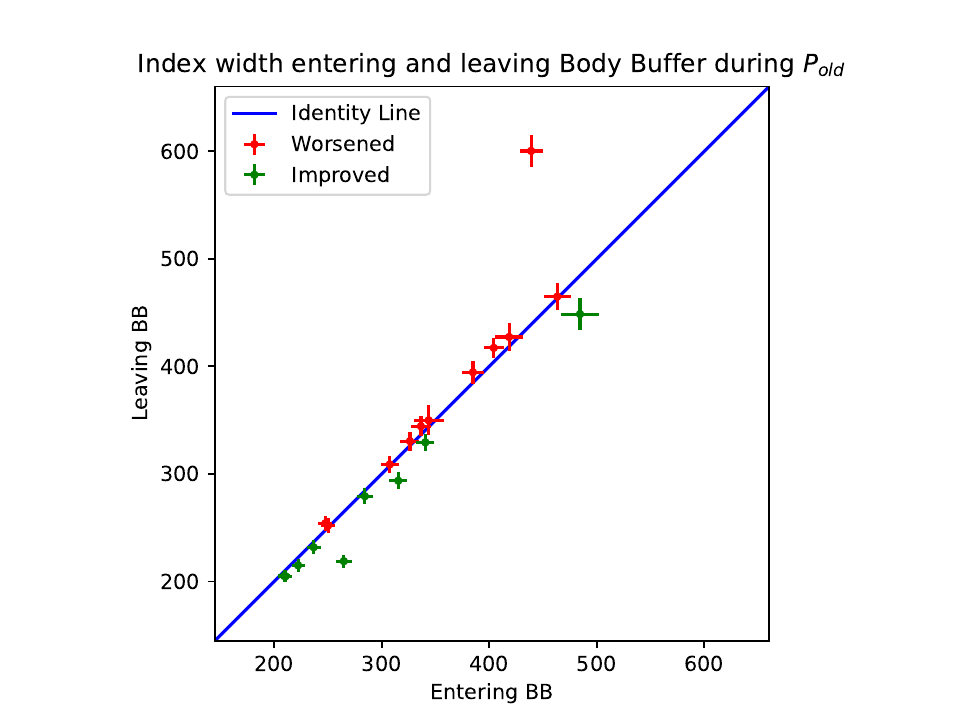}  & \includegraphics[width=0.5\textwidth]{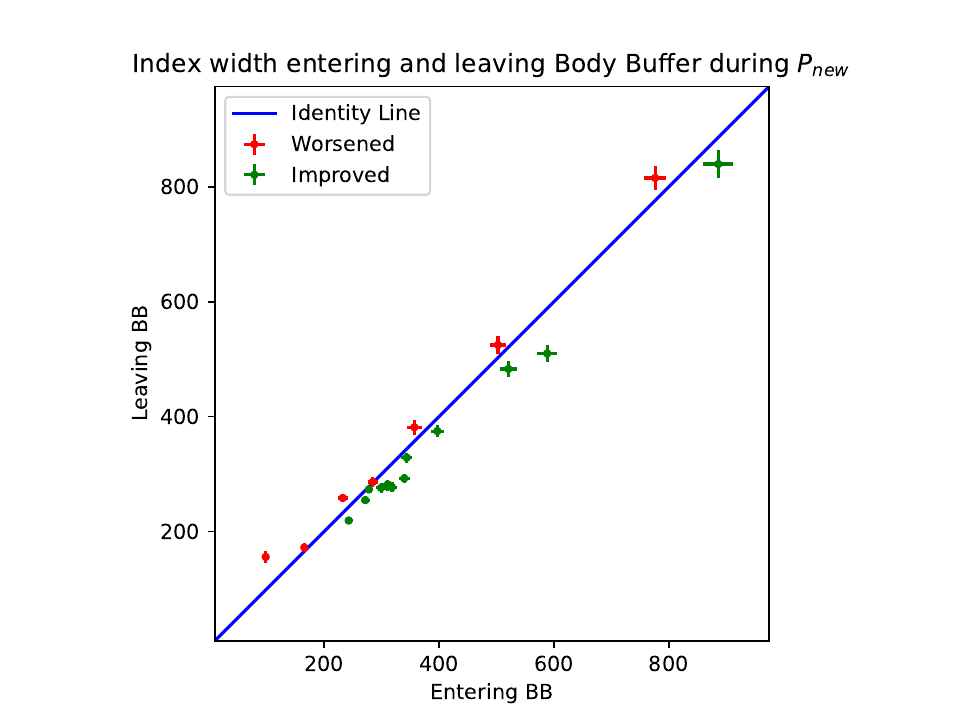}
    \end{tabular}
    \captionof{figure}{Index width entering the \bodybuffer{} ($x$ axis) vs. index width leaving the \bodybuffer{} ($y$ axis) during $P_{old}$ (left) and $P_{new}$ (right). In blue, the identity function $y=x$. The values above the line (red) indicate the days where the index worsened; the values below it (green) indicate the days where it improved.}
    \label{fig:iwp12}
\end{table}

In the left side of Figure~\ref{fig:iwp12} we provide a visualization of the index widths for $P_{old}$. On the $x$ axis, there are the values with respect to the sequence entering the \bodybuffer{}, while on the $y$ axis, there are the values of the respective days leaving the \bodybuffer{}. The blue line represents the identity function ($y=x$). All the points below the line (green) indicate an improvement, while the ones above the line (red) indicate that the index width worsened. In this case, on average, points tend to stay above the line. We can reinforce this claim with a statistical analysis. By performing an orthogonal analysis, the resulting slope is $1.20 \pm 0.09$ with intercept $-58 \pm 26$. The slope is higher than $1$ with a $p$-value of $0.014$. A slope higher than $1$ means that the index widths entering the \bodybuffer{} tends to be greater than those leaving it. In other words, during $P_{old}$ the index width worsened. In the right side of Figure~\ref{fig:iwp12}, we provide the visualization for $P_{new}$. In this case, the points visibly tend to stay below the line. The slope obtained from the orthogonal analysis is $0.90 \pm 0.05$ with intercept $20 \pm 14$. Here, the slope is lower than $1$ with a $p$-value of $0.013$. This value indicates that there is a reduction of the width by \wbdpercent.
\medskip\newline

\section{Conclusion}\label{conclusion}

In this paper, we proposed a multi-objective algorithm for solving the mixed-model assembly line sequencing problem, under consideration of the constraints of the plant, color changeovers in the Paint Shop, and the delivery time of cars -- all simultaneously. We implemented our algorithm as an API and adapted it to the system of a real-world plant, where we collected the data from the actual production process. We showed that our algorithm improved the production performance and quality on all our objectives. With respect to the testing period before the application of our algorithm, we increased the assessed average batch size (aABS) of the sequence leaving the \bodybuffer by 30\%
, which reduced the number of color changeovers in the Paint Shop. We also narrowed the distribution of the cars planned for a specific date, which increases the probability of on-time delivery. We also minimized the distortion of the sequence number distribution. We could not evaluate the impact of our algorithm on the constraint violations, as we did not have access to that data, but we did not receive any significant anomalies or complaints from the plant. We demonstrated the effectiveness and robustness of our algorithm in a real-world setting, which is a rare and valuable contribution in this area of research.

\section{Acknowledgements}\label{acknowledgements}
The research reported in this paper was carried out in the context of the MoDigPro project, which has been supported by the European Regional Development Fund (ERDF). We would like to thank Pranabendu Misra, who co-authored a companion paper with us~\cite{karrenbauer2023improving}, for his valuable contributions and support during this research project.

\section{Disclosure Statement}\label{disclosure}
We made use of a Generative AI tools (Gemini) for generating code for producing figures and for improving language during the first drafts of the Abstract and Introduction.

\section{Data Availability Statement}\label{data availability}\
The authors do not have permission to share data.

\newpage

\bibliographystyle{plain}
\bibliography{ref,local,lib}

\end{document}